\documentclass{elsart}
\usepackage{amsfonts}
\usepackage{graphicx}
\usepackage{amssymb}

\begin{document}

\begin{frontmatter}

\title{Vortex solutions of the discrete
Gross-Pitaevskii equation}
\date{\today}

\author[US]{J Cuevas\thanksref{cor}},
\author[INSA]{G James},
\author[UMA]{PG Kevrekidis},
\author[UMA]{KJH Law}.
\thanks[cor]{Corresponding author.
             E-mail: jcuevas@us.es}

\address[US]{Grupo de F\'{\i}sica No Lineal.
Departamento de F\'{\i}sica Aplicada I. Escuela Universitaria
Polit\'{e}cnica. Universidad de Sevilla. C/ Virgen de \'{A}frica, 7.
41011 Sevilla, Spain}
\address[INSA]{Institut de Math\'ematiques de Toulouse (UMR 5219), INSA de
Toulouse, 135 avenue de Rangueil, 31077 Toulouse Cedex 4, France}
\address[UMA]{Department of Mathematics and Statistics, University of
Massachusetts, Amherst MA 01003-4515, USA}

\journal{Physica D}

\begin{abstract}
In this paper, we consider the dynamical evolution of dark vortex
states in the two-dimensional defocusing discrete nonlinear
Schr{\"o}dinger model, a model of interest both to atomic physics
and to nonlinear optics. We find that in a way reminiscent of their
1d analogs, i.e., of discrete dark solitons, the discrete defocusing
vortices become unstable past a critical coupling strength and, in
the infinite lattice, they apparently remain unstable up to the
continuum limit where they are restabilized. In any finite lattice,
stabilization windows of the structures may be observed. Systematic
tools are offered for the continuation of the states both from the
continuum and, especially, from the anti-continuum limit. Although
the results are mainly geared towards the uniform case, we also
consider the effect of harmonic trapping potentials.
\end{abstract}


\end{frontmatter}

\section{Introduction}

The study of vortices and their existence, stability and dynamical
properties has been a central theme of study in the area of Bose-Einstein
condensates (BECs) \cite{pethick,stringari}. In particular,
the remarkable experiments illustrating the generation of
vortices \cite{vort1,vort2,vort3} and of very robust
lattices thereof \cite{latt1,latt2,latt3} have stirred a tremendous
amount of activity in this area in the past few years, that
has by now been summarized in various reviews and books; see
for example \cite{pismen,fetter,our,book,berloff1,ournew}. Much of this
activity has been centered around the robustness of vortex
structures in the context of the mean-field dynamics of the
BECs (which are controllably accurately described by a nonlinear
Schr{\"o}dinger (NLS) equation) in the presence of
many of the potentials that are relevant to the
trapping of atomic BECs including parabolic traps \cite{pethick,stringari}
and periodic optical lattice ones \cite{konotop,morsch}.
Particularly, the latter context of optical lattice potentials
is quite interesting, as it has been suggested that vortices
(for example of topological charge $S=1$) will be unstable
when centered at a minimum of the lattice potential \cite{jpb},
an instability that it would be interesting to understand in more detail.

On the other hand, the BECs in the presence of periodic potentials
have been argued to be well-approximated by models of the
discrete nonlinear Schr{\"o}dinger (DNLS) type (i.e., resembling the
finite-difference discretization of the continuum equation)
\cite{TS,konotop2,us,usnew}. In that regard, to understand the
existence and stability properties of vortices in the presence of
periodic potentials, it would be interesting to analyze the
discrete analog of the relevant NLS equation. This is also interesting
from a different perspective in this BEC context, namely that if
finite-difference schemes are employed to analyze the properties of
the continuum equation, it is useful to be aware of features introduced
by virtue of the discretization.

However, it should be stressed that this is not a problem of
restricted importance in the context of quantum fluids; it is also
of particular interest in nonlinear optics where two-dimensional
optical waveguide arrays have been recently systematically
constructed e.g. in fused silica in the form of square lattices
\cite{lederer1,lederer2} (and, more recently of even more complex
hexagonal lattices \cite{lederer3}), whereby discrete solitons can
be excited. By analogy to their one-dimensional counterparts of
discrete dark solitons, which have been created in defocusing
waveguide arrays with the photovoltaic nonlinearity \cite{kip}, we
expect that it should be possible to excite discrete dark vortices
in defocusing two-dimensional waveguide arrays. An especially
interesting feature of dark solitons that was observed initially in
\cite{johansson} (see also \cite{fitrakis}) is that on-site discrete
dark solitons are stable for sufficiently coarse lattices, but they
become destabilized beyond a certain coupling strength among
adjacent lattice sites and remain so until the continuum limit where
they are again restabilized (as the point spectrum eigenvalue that
contributes to the instability becomes zero due to the restoration
of the translational invariance in the continuum problem)
\cite{johansson,fitrakis}. It is therefore of interest to examine if
the instability mechanisms of discrete defocusing vortices are of
this same type or are potentially different and how the relevant
stability picture is modified as a function of the inter-site
coupling strength.

It is this problem of the existence, stability and continuation of
the vortex structures as a function of coupling strength that
we examine in the present work.
We consider, in particular, a two-dimensional
discrete non-linear Schr\"odinger equation
\begin{equation}\label{eq:dyn}
    i\frac{d\psi_{n,m}}{dt}-|\psi_{n,m}|^2\psi_{n,m}+\epsilon \Delta \psi_{n,m}=0,
\end{equation}
where $\Delta \psi_{n,m}=
\psi_{n+1,m}+\psi_{n-1,m}+\psi_{n,m+1}+\psi_{n,m-1}-4 \psi_{n,m}$ is
the discrete Laplacian. We study the {\em defocusing} case when
$\epsilon >0$. In that case, equation (\ref{eq:dyn}) is denoted as
discrete Gross-Pitaevskii equation in analogy with its continuum
counterpart \cite{pethick,stringari,Berloff}.

We look for time-periodic solutions with frequency $\omega$. Using
the ansatz $\psi_{n,m}(t)= \sqrt{\omega}\, \phi_{n,m} e^{-i \omega
t}$, we obtain
\begin{equation}
\label{eq:stat}
C\, \Delta \phi_{n,m} + (1-|\phi_{n,m}|^2) \phi_{n,m}=0,
\end{equation}
where we have set $C = \epsilon / \omega$. The coupling parameter
$C>0$ determines the strength of discreteness effects. The limit $C
\rightarrow +\infty$ corresponds to the continuum (stationary)
Gross-Pitaevskii equation:
\begin{equation}
\label{eq:statcont} \frac{\partial^2\phi}{\partial
x^2}+\frac{\partial^2\phi}{\partial y^2}+ (1-|\phi |^2) \phi=0.
\end{equation}

The case $C \rightarrow 0$ corresponds to the so-called
anti-continuum (AC) limit \cite{MA94}.

When equation (\ref{eq:stat}) is considered on an infinite lattice
$\mathbb{Z}^2$, we look for solutions satisfying
$|\phi_{n,m}|\rightarrow1$ when $(n,m)\rightarrow\infty$, for which
$\phi_{n,m}$ vanishes at one lattice site, e.g. at $(n,m)=(0,0)$.
Such solutions are denoted as discrete vortices, or ``dark'' vortex
solitons. If one trigonometric turn on any path
$\mbox{Max}(|n|,|m|)=\rho$ around the vortex center changes the
argument of $\phi_{n,m}$ by $2 \pi S$ ($S\in \mathbb{Z}$), then the
vortex is said to have a topological charge (or vorticity) equal to
$S$.

In this paper we numerically investigate the existence and stability
of such solutions on a finite lattice of size $N\times N$, $N$ being
large; our analysis is performed as a function of the lattice
coupling parameter $C$ and we illustrate how to perform relevant
continuations both from the continuum, as well as, more importantly
from the AC limit (section 2). We mainly focus on numerically
computing vortex solutions with vorticity $S=1$ and $S=2$ (section
3). In section 4, we also obtain such solutions in the presence
of an external harmonic trap (the latter is typically present in BEC
experiments).
Finally, section 5 presents our conclusions and some future
directions of potential interest.

\section{Numerical method}
\label{num_method}

We compute vortex solutions of (\ref{eq:stat}) using the Newton method
and a continuation with respect to $C$.
The path-following can be initiated either near the continuum limit
(for $C$ large) or at the anti-continuum limit $C=0$, since in both cases
one is able to construct a suitable initial guess for the Newton method.

For relatively high $C$, a suitable initial condition
for a vortex with topological charge $S$ is obtained with
a Pad\'e approximation developed for the
continuum limit in \cite{Berloff}. We set
$\phi_{n,m}=\rho_{n,m}e^{iS\alpha_{n,m}}$, where
\begin{equation}
    \rho_{n,m}=\sqrt{\frac{r_{n,m}^{2S}(a_1+a_2r_{n,m}^2)}
    {1+b_1r_{n,m}^2+a_2r_{n,m}^{2S+2}}},
    \ \ \ \  r_{n,m}=\sqrt{n^2+m^2}
\end{equation}
($a_1=11/32$, $a_2=a_1/12$, $b_1=1/3$, see reference \cite{Berloff}),
$$
\alpha_{n,m}=
\left\{
\begin{array}{ll}
\arctan(m/n)  + \frac{3\pi}{2} \,      &\mbox{ for } n\geq 1,    \\
\arctan(m/n) + \frac{\pi}{2} \,       &\mbox{ for } n\leq -1,    \\
\frac{\pi}{2}\, (1-\mbox{sign}(m)) &\mbox{ for } n=0.
\end{array}
\right.
$$
Once a vortex is found for a given $C$, the solution can be
continued by increasing or decreasing $C$.
Although this method was found to be efficient, it remains limited to
single vortex solutions having explicit continuum approximations.
Moreover, when the Newton method is applied to continue
these solutions near $C=0$, the Jacobian matrix becomes
ill-conditioned (and non-invertible for $C=0$)
and the iteration does not converge.

In what follows we introduce a different method
having a wider applicability, and for which the
above mentioned singularity is removed.
We consider a finite $N\times N$ lattice with
$(n,m) \in \Gamma = \{-M,\ldots ,M\}^2$
($N=2M+1$), equipped with
fixed-end boundary conditions given below.
We set $\phi_{n,m} = R_{n,m} \, e^{i \theta_{n,m} }$ and note
$R = (R_{n,m})_{n,m}$, $\theta = (\theta_{n,m})_{n,m}$.
One obtains the equivalent problem
\begin{equation}
\label{acm2}
R_{n,m}\,  (1- R_{n,m}^2)+ C \, f(R,\theta )_{n,m} =0,
\end{equation}
\begin{equation}
\label{acp}
C \, g(R,\theta )_{n,m}=0,
\end{equation}
where $f(R,\theta ) = \mbox{Re }[\, e^{-i\theta }\, \Delta (R\,
e^{i\theta } )\, ]$ and $g(R,\theta ) = \mbox{Im }[\, e^{-i\theta
}\, \Delta (R\, e^{i\theta } )\, ]$ can be rewritten \vspace{1cm}
\begin{eqnarray*} f(R,\theta )_{n,m} & = & R_{n+1,m}
\cos{(\theta_{n+1,m}-\theta_{n,m})} + R_{n-1,m}
\cos{(\theta_{n,m}-\theta_{n-1,m})} -4 R_{n,m}
\\
 &  & +R_{n,m+1} \cos{(\theta_{n,m+1}-\theta_{n,m})}
 + R_{n,m-1} \cos{(\theta_{n,m}-\theta_{n,m-1})},
 \end{eqnarray*}
\begin{eqnarray*}
g(R,\theta )_{n,m} & = &
R_{n+1,m} \sin{(\theta_{n+1,m}-\theta_{n,m})}
- R_{n-1,m} \sin{(\theta_{n,m}-\theta_{n-1,m})}
\\
 &  & +R_{n,m+1} \sin{(\theta_{n,m+1}-\theta_{n,m})}
 - R_{n,m-1} \sin{(\theta_{n,m}-\theta_{n,m-1})}.
 \end{eqnarray*}
 Now we divide equation (\ref{acp}) by $C$
 (this eliminates the above-mentioned degeneracy at $C=0$)
 and consider equation (\ref{acm2}) coupled to
\begin{equation}
\label{acp2}
g(R,\theta )_{n,m}=0.
\end{equation}
System (\ref{acm2}), (\ref{acp2}) is supplemented by the boundary conditions
\begin{eqnarray}
\label{rinf}
 R_{n,m} =  1 &    \mbox{ for }& \mbox{Max}(|n|,|m|)=M, \\
 \label{tinf}
\theta_{n,m} = \theta_{n,m}^\infty
&    \mbox{ for }& \mbox{Max}(|n|,|m|)=M.
\end{eqnarray}
The prescribed value $\theta_{n,m}^\infty$ of the angles on the
boundary will depend on the type of vortex solution we look for. In
particular, we use the boundary conditions $\theta_{n,m}^\infty = S
\alpha_{n,m}$ for a single vortex with topological charge $S$
centered at $(n,m)=(0,0)$.

For $C=0$, a single vortex at $(n,m)=(0,0)$ corresponds to fixing
$R_{0,0}=0$ and $R_{n,m}=1$ everywhere else. Equation (\ref{acp2})
yields in that case
\begin{equation}
\label{aclp1}
\begin{array}{l}
\sin{(\theta_{n+1,m}-\theta_{n,m})}-\sin{(\theta_{n,m}-\theta_{n-1,m})} \\
+\sin{(\theta_{n,m+1}-\theta_{n,m})}-\sin{(\theta_{n,m}-\theta_{n,m-1})}
=0,
 \\
(n,m)\in \Gamma \setminus \{ \, (0,\pm 1),\, (\pm 1,0)\, ,\, (\pm M, m)\, ,\, (n, \pm M)\, \}
\end{array}
\end{equation}
supplemented by the four following relations at
$(n,m)= (0,\pm 1),\, (\pm 1,0)$
\begin{eqnarray}
\label{aclp2}
\sin{(\theta_{1,\pm 1}-\theta_{0, \pm 1})}-\sin{(\theta_{0,\pm 1}-\theta_{-1,\pm 1})}
+\sin{(\theta_{0,\pm 2}-\theta_{0,\pm 1})} &=&0, \\
\label{aclp3}
\sin{(\theta_{\pm 2,0}-\theta_{\pm 1,0})}+\sin{(\theta_{\pm 1,1}-\theta_{\pm 1,0})}
-\sin{(\theta_{\pm 1,0}-\theta_{\pm 1,-1})} &=&0.
\end{eqnarray}
For a vortex with topological charge $S=1$,
solutions of (\ref{tinf})-(\ref{aclp3}) are
computed by the Newton method, starting from the initial guess
$\theta_{n,m} = \alpha_{n,m}$.
The symmetries of the problem allow one to divide by four the size of the computational domain. Indeed one can take $(n,m) \in  \{ 0,\ldots ,M\}^2$ with the boundary conditions $\theta_{0,m} = \alpha_{0,m}$,  $\theta_{n,0} = \alpha_{n,0}$. Solutions on the whole lattice $\Gamma$ have the symmetries
\begin{equation}
\label{symtheta}
\theta_{n,-m} = \pi- \theta_{n,m} \, [2\pi], \ \ \ \
\theta_{-n,m} = - \theta_{n,m}\, [2\pi] .
\end{equation}
These conditions make (\ref{aclp1}) automatically satisfied at
$(n,m)=(0,0)$ ($\theta_{0,0}$ need not being specified). Afterwards,
the corresponding solution of (\ref{acm2}),
(\ref{acp2})-(\ref{tinf}) can be continued to $C > 0$ by the Newton
method, yielding a solution $\phi_{n,m}=R_{n,m}e^{i\theta_{n,m}}$ of
(\ref{eq:stat}) (see section \ref{numres}). For higher topological
charges, the initial guess
$\tilde\phi_{n,m}=R_{n,m}e^{iS\theta_{n,m}}$ can be used to compute
a vortex solution of (\ref{eq:stat}) by the Newton method. This is
done in section \ref{numres} also for $S=2$. All these continuations
are performed with a $10^{-8}$ accuracy.

\section{\label{numres}Numerical computation of single vortices}

In this section we analyze the existence and stability of discrete
vortices centered on a single site, as a function of the coupling
strength $C$ for fixed-end boundary conditions. The stability of the
discrete vortex solitons is studied assuming small perturbations in
the form of $\delta\psi_{m,n}=\exp(-i t)[p_{n,m}\exp(-i\lambda
t)+q_{n,m}\exp(i\lambda^* t)]$, the onset of instability indicated
by the emergence of $\mathrm{Im}(\lambda)\neq 0$; $\lambda$ in this
setting denotes the perturbation eigenfrequency. Note that it is
sufficient to consider the case $\omega=1$ for stability
computations, because this case can always be recovered by rescaling
time.

\begin{figure}
\begin{center}
\begin{tabular}{cc}
    \includegraphics[width=7cm]{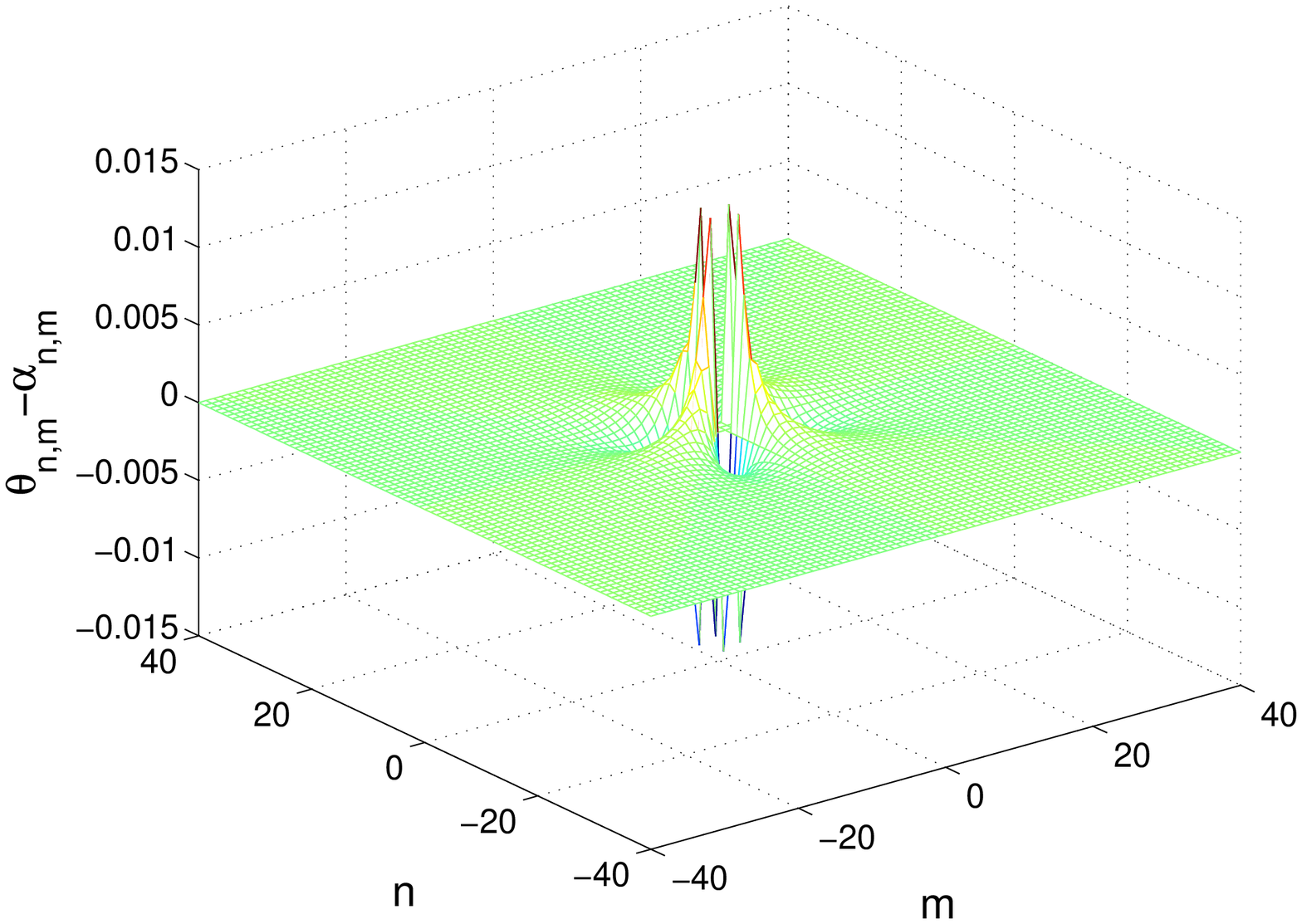} &
    \includegraphics[width=7cm]{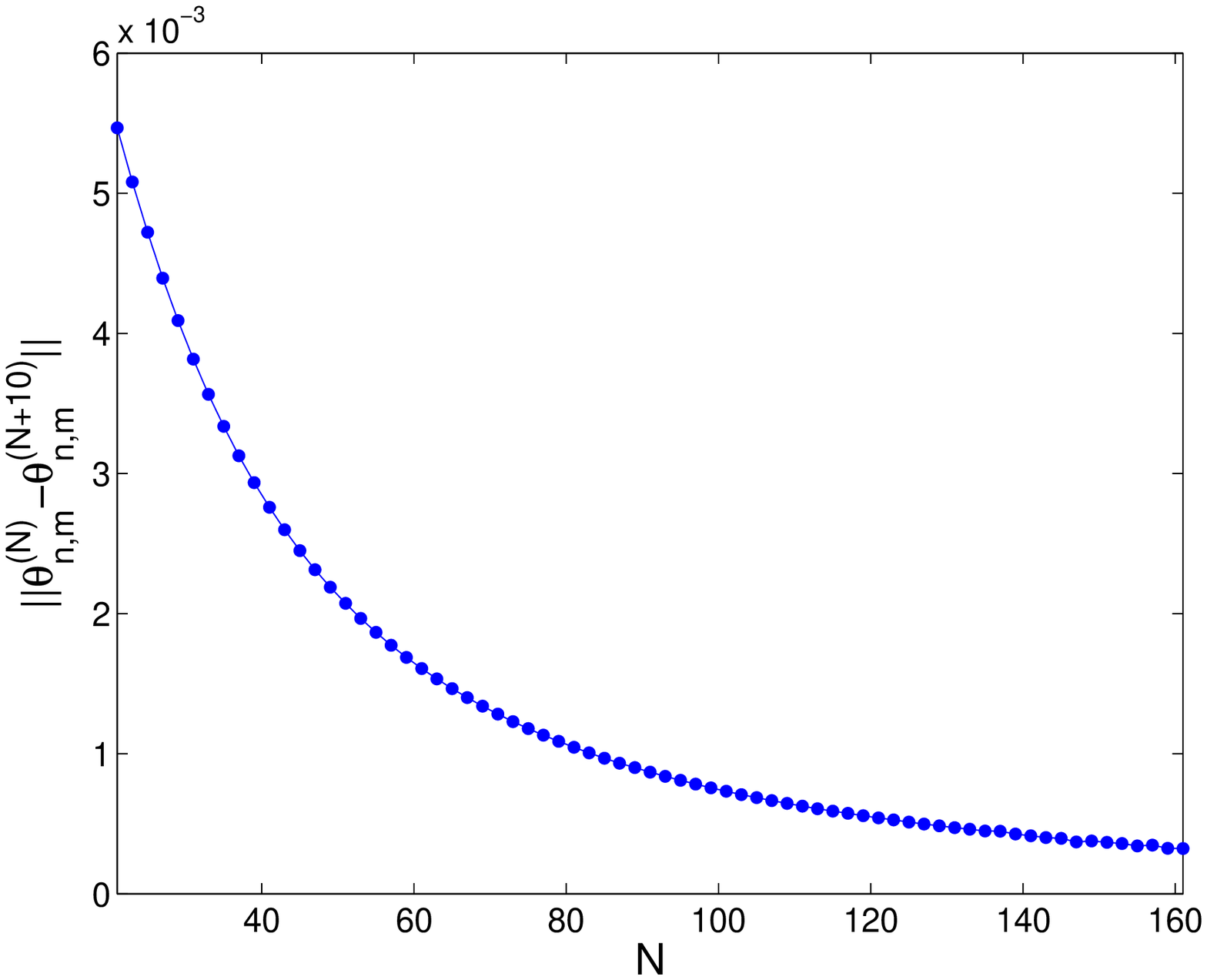}
\end{tabular}
\caption{(Left panel) The spatial profile of the difference between
the computed angles and the seed
angles in a $81\times81$ lattice at the AC-limit. (Right panel)
Dependence of $||\theta_{n,m}^{N}-\theta_{n,m}^{N+10}||_{\infty}$ with
respect to the lattice size $N$. In both cases, the lattice has
fixed end boundary conditions.}
\label{fig:AC1}
\end{center}
\end{figure}

Figure \ref{fig:AC1} compares the computed angles $\theta_{n,m}$
with respect to the seed angle $\alpha_{n,m}$ for fixed-end boundary
conditions and $N=81$. The most significant differences arise close
to the vortex center. This figure also shows the dependence on $N$
of the difference between the angles $\theta$ for a given domain
size $N$ and for a larger domain of size $N+10$. This is done
through $||\theta_{n,m}^{N}-\theta_{n,m}^{N+10}||$ where $||\cdot||$
is the $\infty$-norm, and $\theta_{n,m}^{N}$ represent the angles at
a given lattice size $N$. The main contribution of this norm
corresponds to the boundary sites. On the other hand, the decrease
of this norm as a function of $N$ originates from the convergence of
the configuration to an asymptotic form.

\begin{figure}
\begin{center}
    \includegraphics[width=7cm]{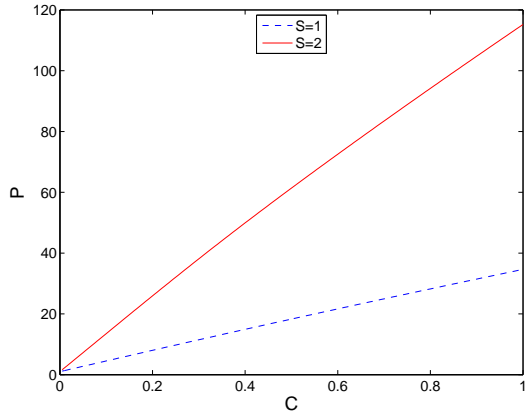}
\caption{Dependence of the complementary norm on the coupling strength $C$
for $S=1$ and $S=2$.} \label{fig:power}
\end{center}
\end{figure}

Figure \ref{fig:power} shows the complementary norm of the
$S=1$ and $S=2$ vortices, which is defined as \cite{MHM07}:

\begin{equation}
    P=\sum_n\sum_m\left(|\phi_{\infty}|^2-|\phi_{n,m}|^2\right)
\end{equation}

with $|\phi_{\infty}|^2$ being the background density; in our case,
$|\phi_{\infty}|^2=1$. As it can be observed in the figure, vortices
with $S=1$ and $S=2$ can be continued for couplings up to O$(1)$ and
presumably for all $C$\footnote{In fact, vortices have been
continued at least up to $C=10$ without any convergence problems,
and their existence in the continuum limit suggests that it should
be, in principle, possible to identify such structures for
arbitrarily large values of $C$.}. It should be mentioned in passing
that the method has also been successfully used to perform
continuation in the vicinity of the anti-continuum limit, even for
higher charge vortices such as $S=3$. Notice also that all the
considered solutions are ``black'' solitons, i.e., the vortex center
has amplitude $R_{0,0}=0$.

\begin{figure}
\begin{center}
\begin{tabular}{cc}
    \includegraphics[width=7cm]{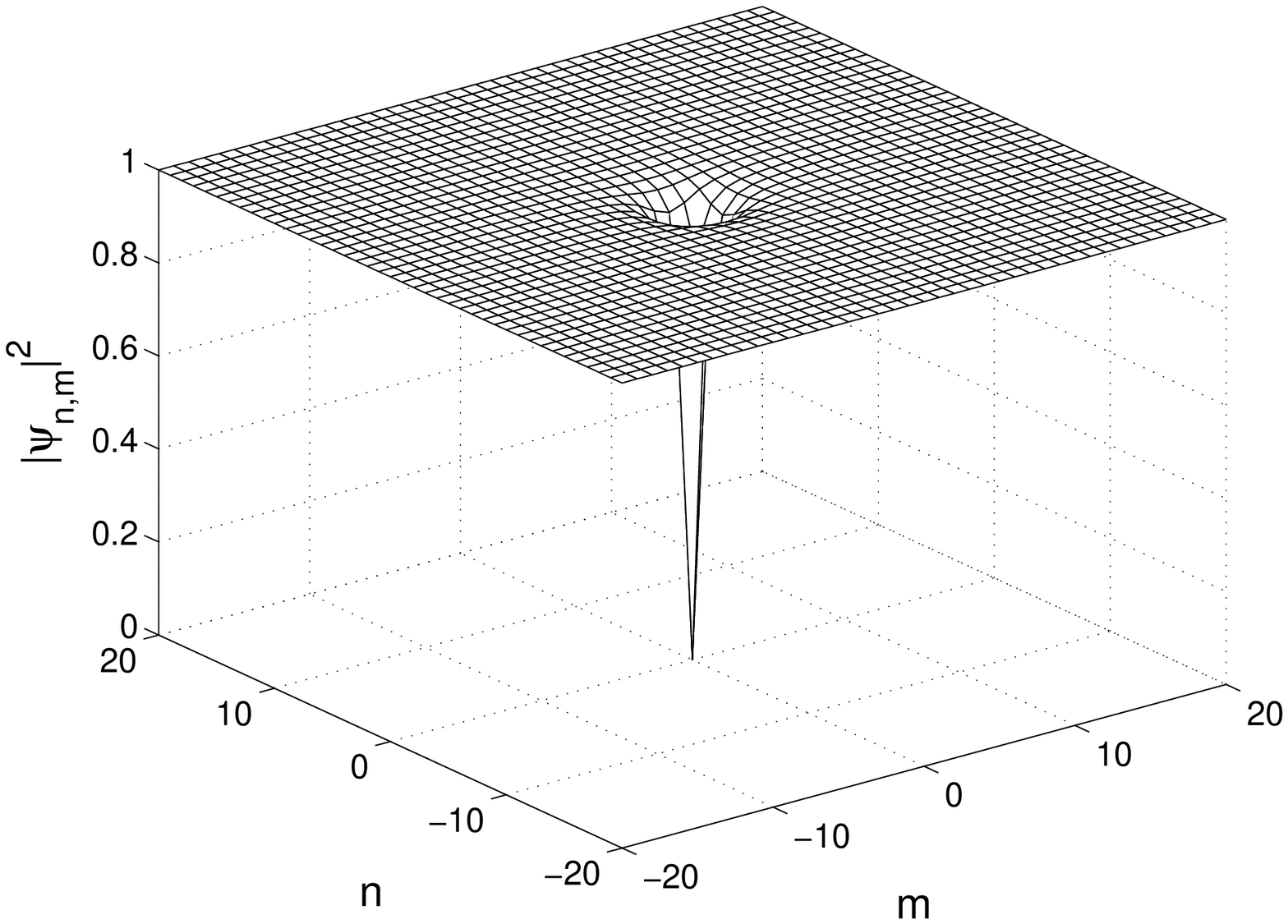} &
    \includegraphics[width=7cm]{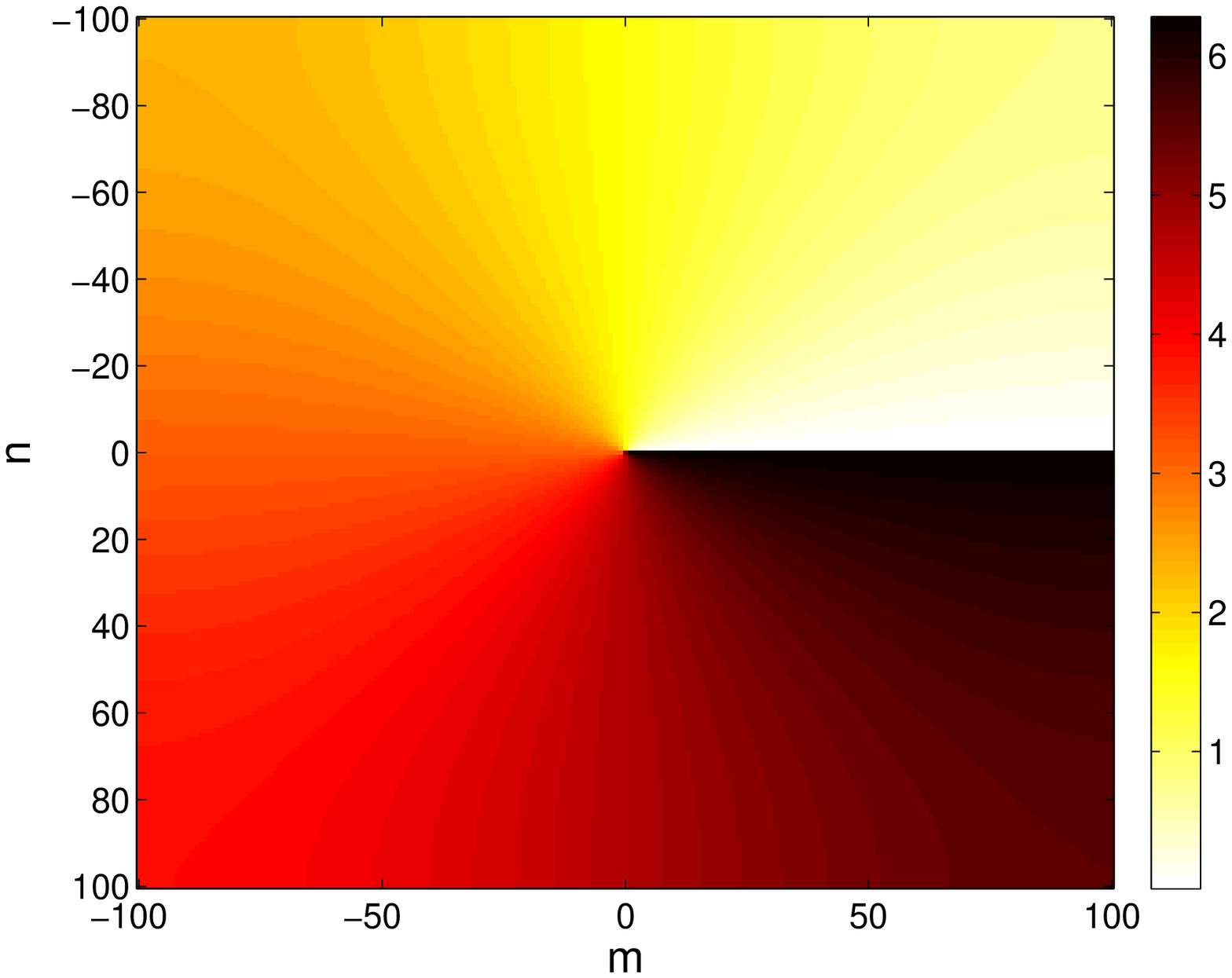} \\
    \includegraphics[width=7cm]{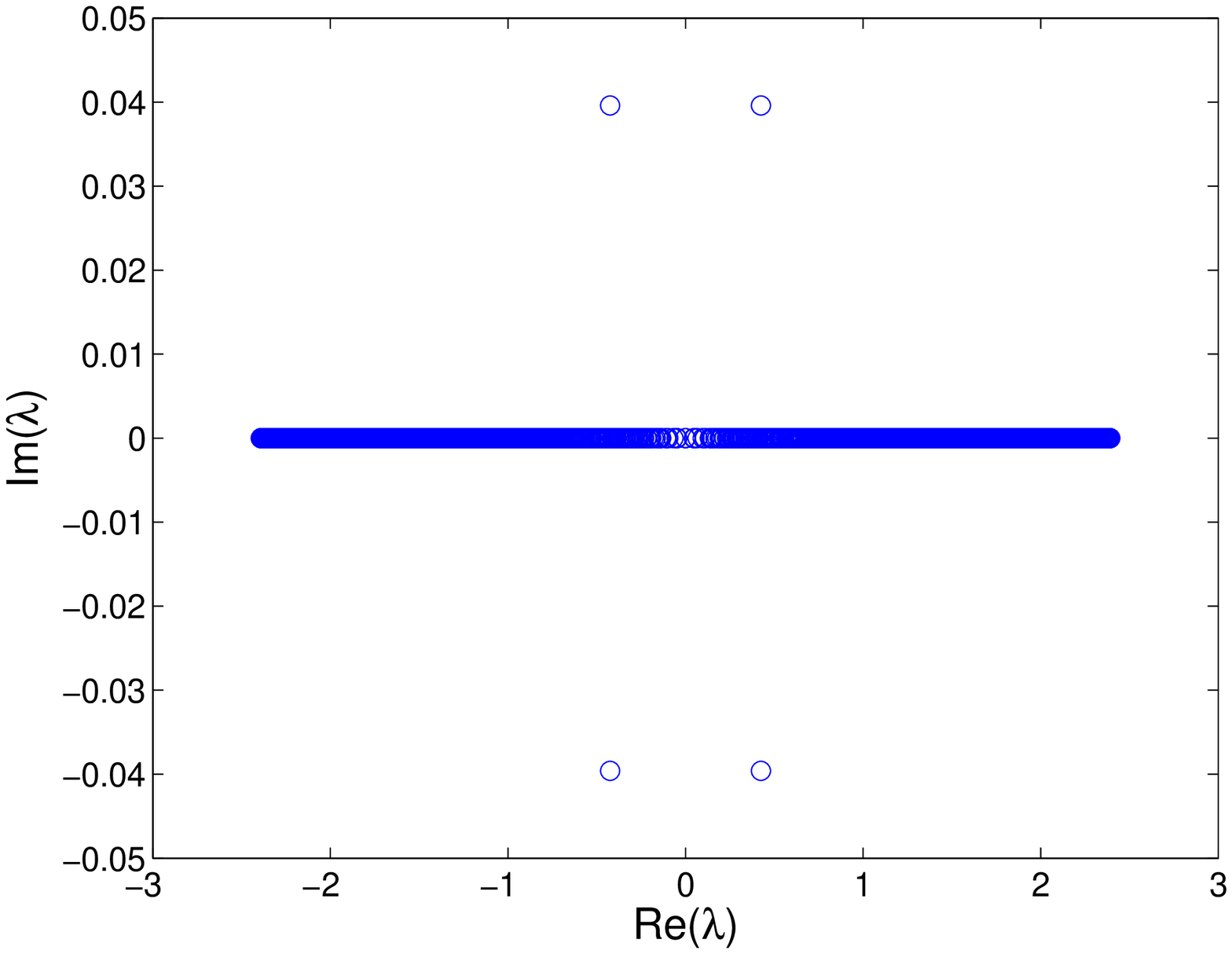} &
    \includegraphics[width=7cm]{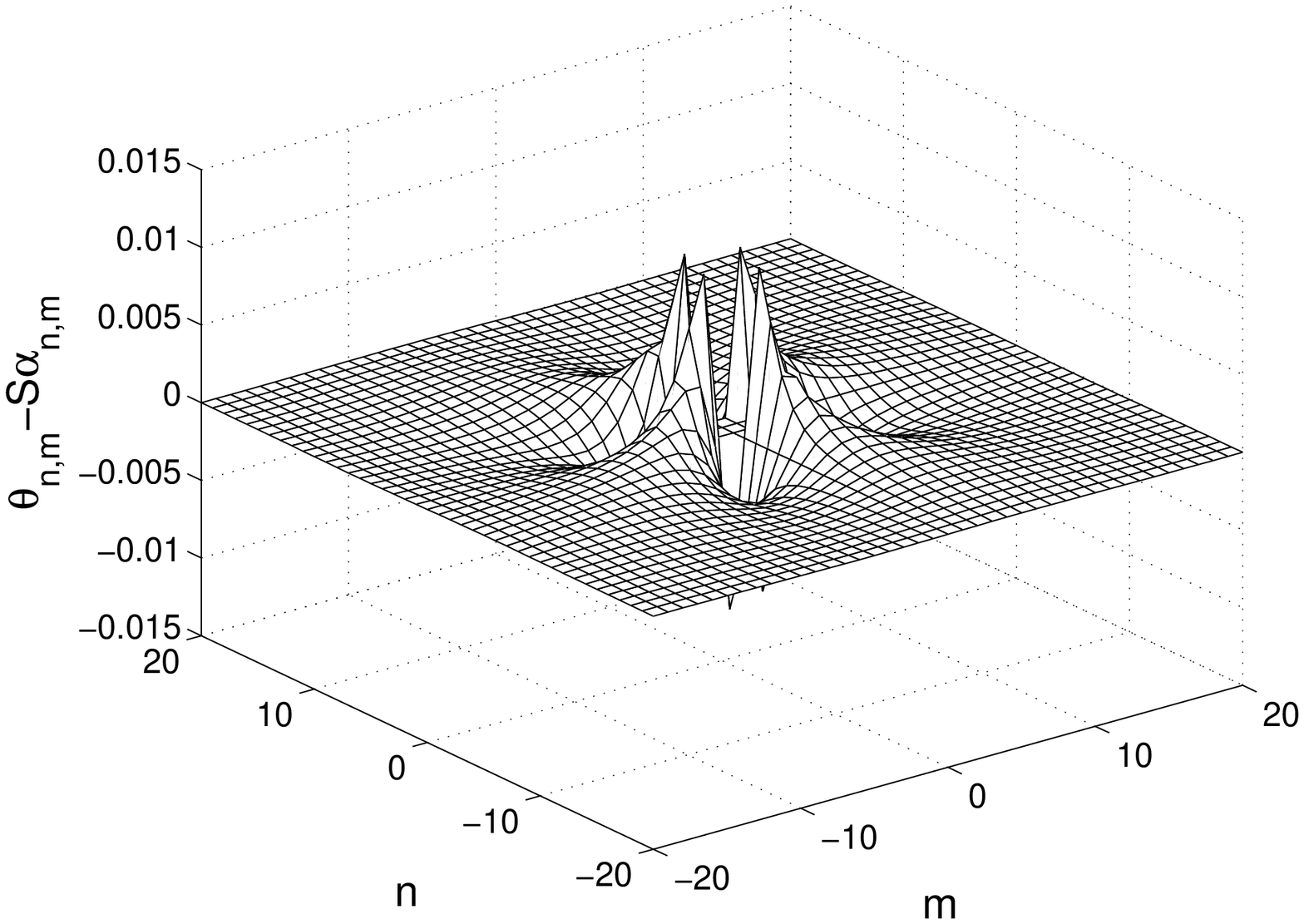} \\
\end{tabular}
\caption{Vortex soliton with $S=1$ and $C=0.2$. (Top left panel) density
Profile; (top right panel) angular dependence; (bottom left panel)
spectral plane of stability eigenfrequencies [recall that the presence
of eigenfrequencies with non-vanishing imaginary part denotes instability];
(bottom right panel) comparison of the vortex angles with
$\alpha_{n,m}$.} \label{fig:S1}
\end{center}
\end{figure}

\begin{figure}
\begin{center}
\begin{tabular}{cc}
    \includegraphics[width=7cm]{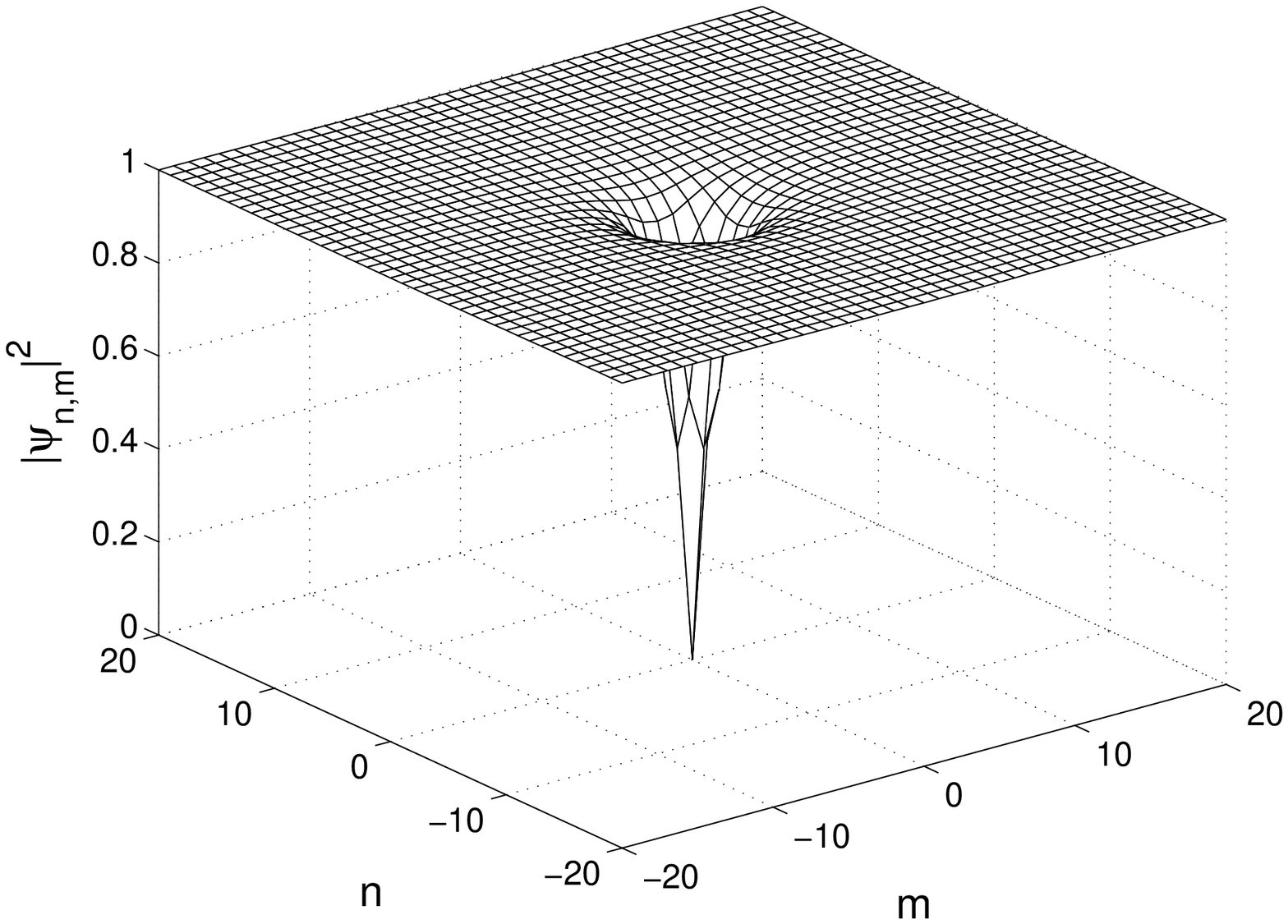} &
    \includegraphics[width=7cm]{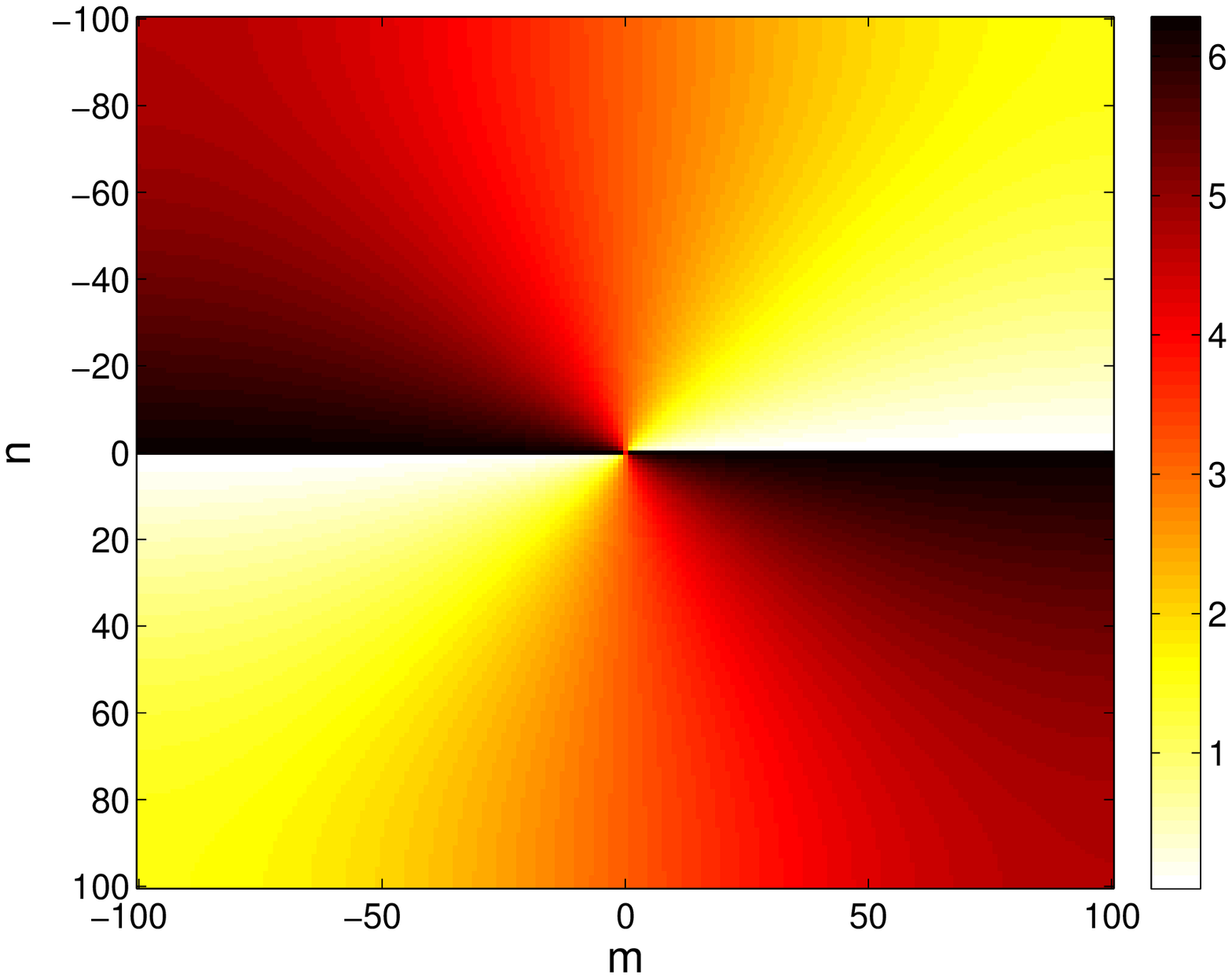} \\
    \includegraphics[width=7cm]{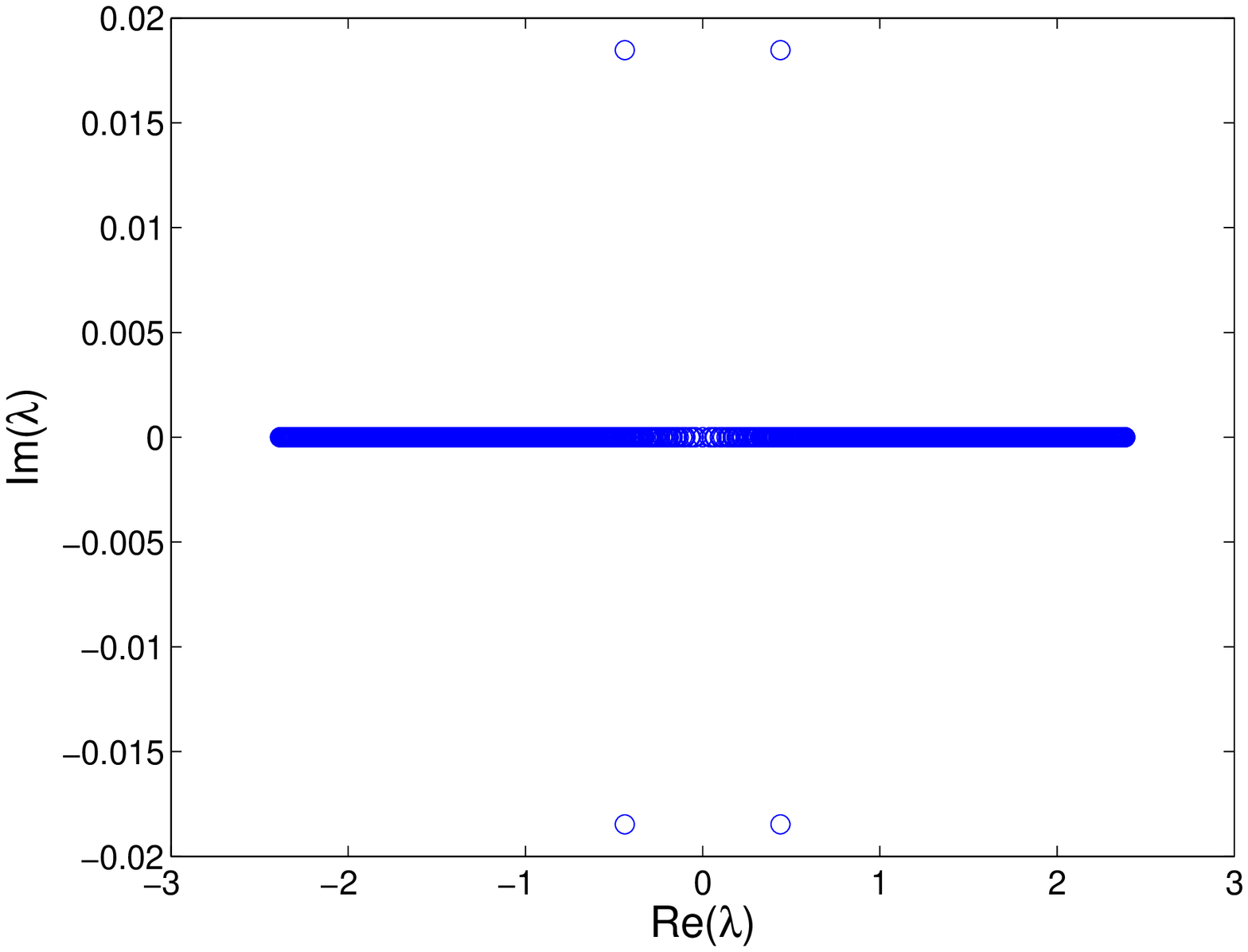} &
    \includegraphics[width=7cm]{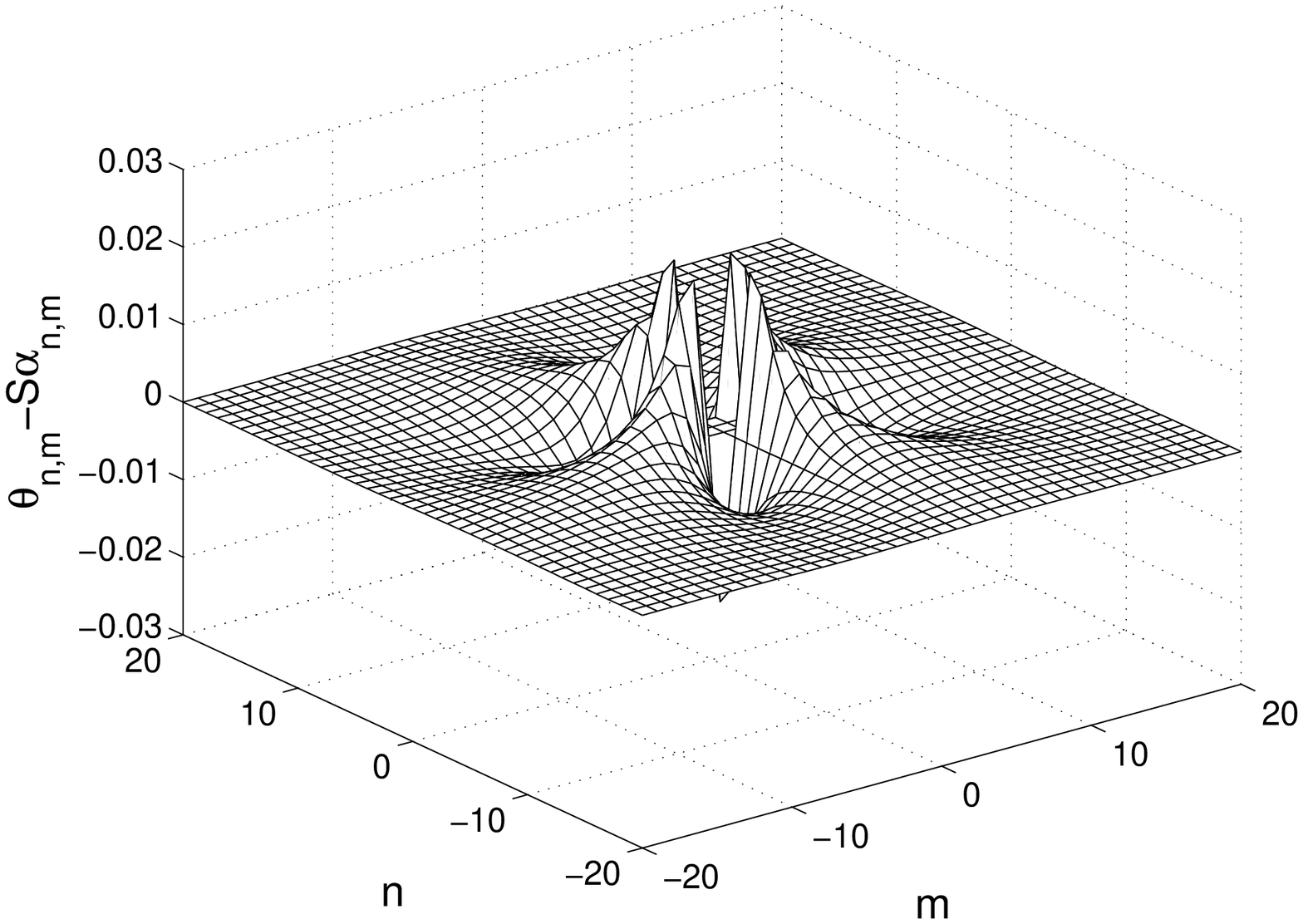} \\
\end{tabular}
\caption{Same as Fig. \ref{fig:S1} but for $S=2$.} \label{fig:S2}
\end{center}
\end{figure}

Figures \ref{fig:S1} and \ref{fig:S2}  show, for $S=1$ and $S=2$
vortices, respectively, the profile
$|\psi_{n,m}|^2=|\phi_{n,m}|^2=R_{n,m}^2$, the angles
$\theta_{n,m}$, the spectral plane of the stability eigenfrequencies
and a comparison with the angles $\alpha_{n,m}$. In all cases,
$C=0.2$ is shown, which corresponds to unstable vortices.

\begin{figure}
\begin{center}
\begin{tabular}{cc}
    \includegraphics[width=7cm]{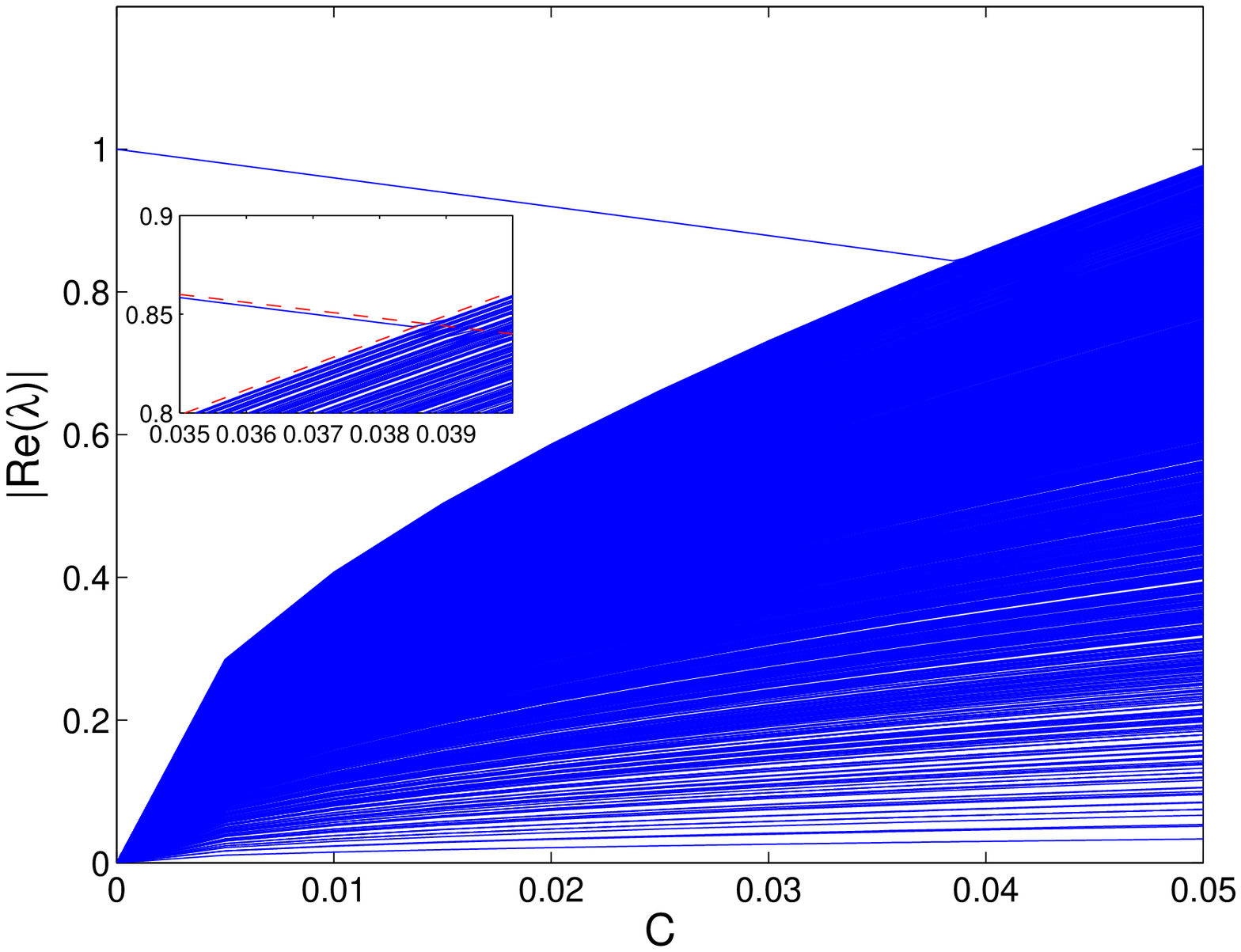} &
    \includegraphics[width=7cm]{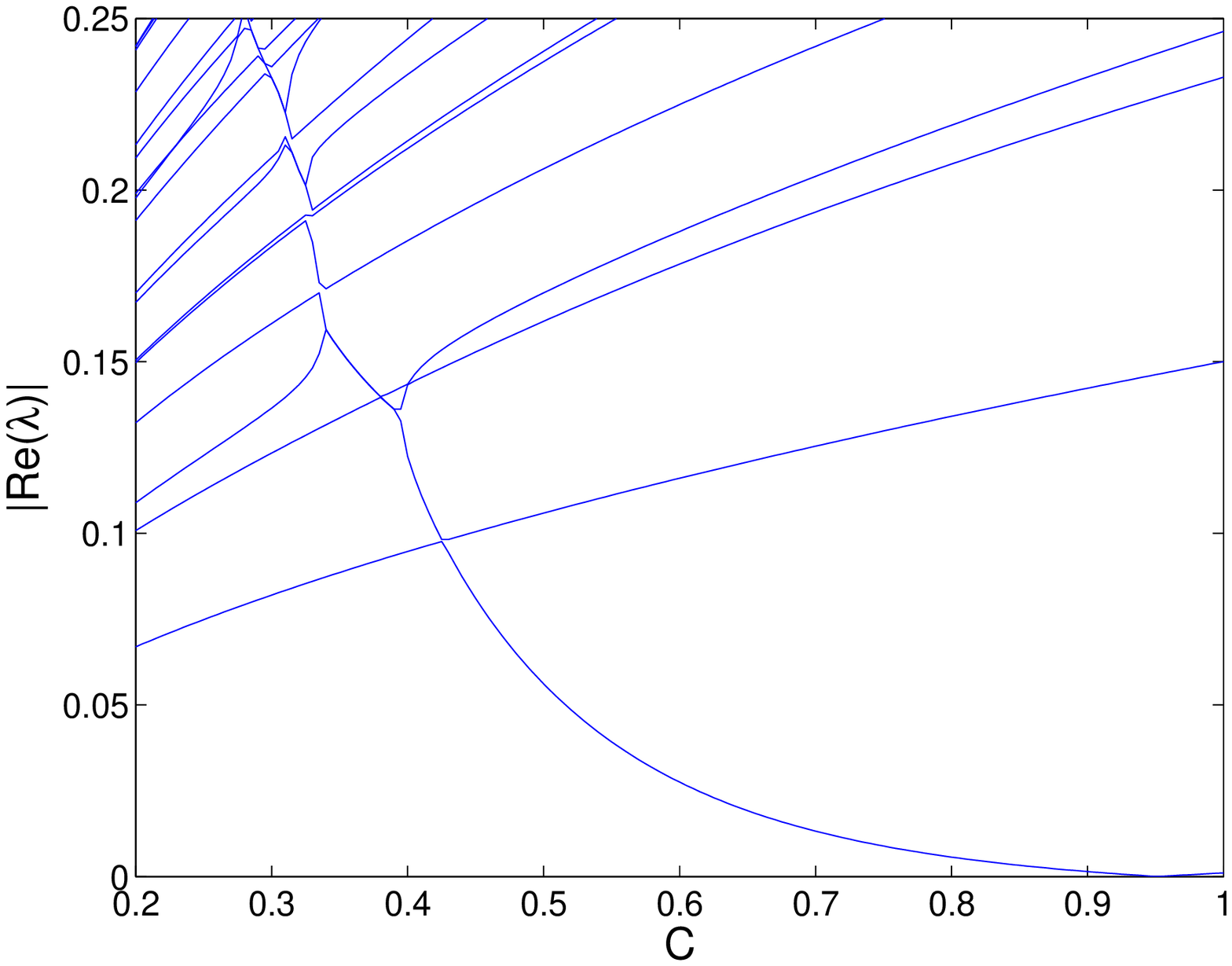}
\end{tabular}
\caption{Real part of the stability eigenfrequencies for $S=1$. The
panels show zooms of two different regions. Dashed lines correspond
to the predicted eigenvalues $\lambda \approx 1-4 C$ and $\lambda
\approx\sqrt{64 C^2 + 16 C}$.} \label{fig:S1stab}
\end{center}
\end{figure}

The vortices with $S=1$ and $S=2$ are, respectively, stable for
$C<C_{cr}\approx 0.0395$ and $C<C_{cr}\approx0.0425$. This
instability, highlighted in the case of the $S=1$ vortex in Fig.
\ref{fig:S1stab} can be rationalized by analogy with the
corresponding stability calculations in the case of dark solitons
\cite{johansson}. In particular, the relevant linearization problem
can be written in the form:
\begin{eqnarray}
\lambda \left( \begin{array}{c} p_{n,m} \\ q_{n,m}^{\star}
\end{array} \right) = \left( \begin{array}{cc}
2  |\phi_{n,m}|^2 -1 - C \Delta &  \phi_{n,m}^2 \\
-(\phi_{n,m}^2)^{\star} & 1 -2 |\phi_{n,m}|^2 + C \Delta
\end{array} \right) \left( \begin{array}{c} p_{n,m} \\ q_{n,m}^{\star}
\end{array} \right).
\label{ceq4b}
\end{eqnarray}
However, by analogy to the corresponding 1d problem, the symmetry
and the high spatial localization of the localized eigenvector at
low coupling renders it a good approximation to write for the
relevant perturbations that $\Delta p_{n,m} \approx -4 p_{n,m}$ (and
similarly for $q$), by virtue of which it can be extracted that the
relevant eigenfrequency is $\lambda \approx 1-4 C$.
This leading order prediction (as a function of $C$)
for the internal (``translational'')
mode frequency is based on the anti-symmetry of both the real and
the imaginary parts of the vortex configuration around its central site,
in analogy with the anti-symmetry of the on-site dark soliton around
its central site in the 1d analog of the problem \cite{johansson,fitrakis}.
This feature (whose continuation to the $C \rightarrow \infty$ leads to
a zero frequency mode due to the translational invariance of the
underlying continuum model) is an example of the
``negative energy'' modes that both dark solitons (see e.g.,
\cite{giorgo} and references therein) and vortices (see e.g. \cite{pu})
are well-known to possess (due to the fact that, although stationary,
they are not ground states of the respective 1d and 2d systems).

\begin{figure}
\begin{center}
\begin{tabular}{cc}
    \includegraphics[width=7cm]{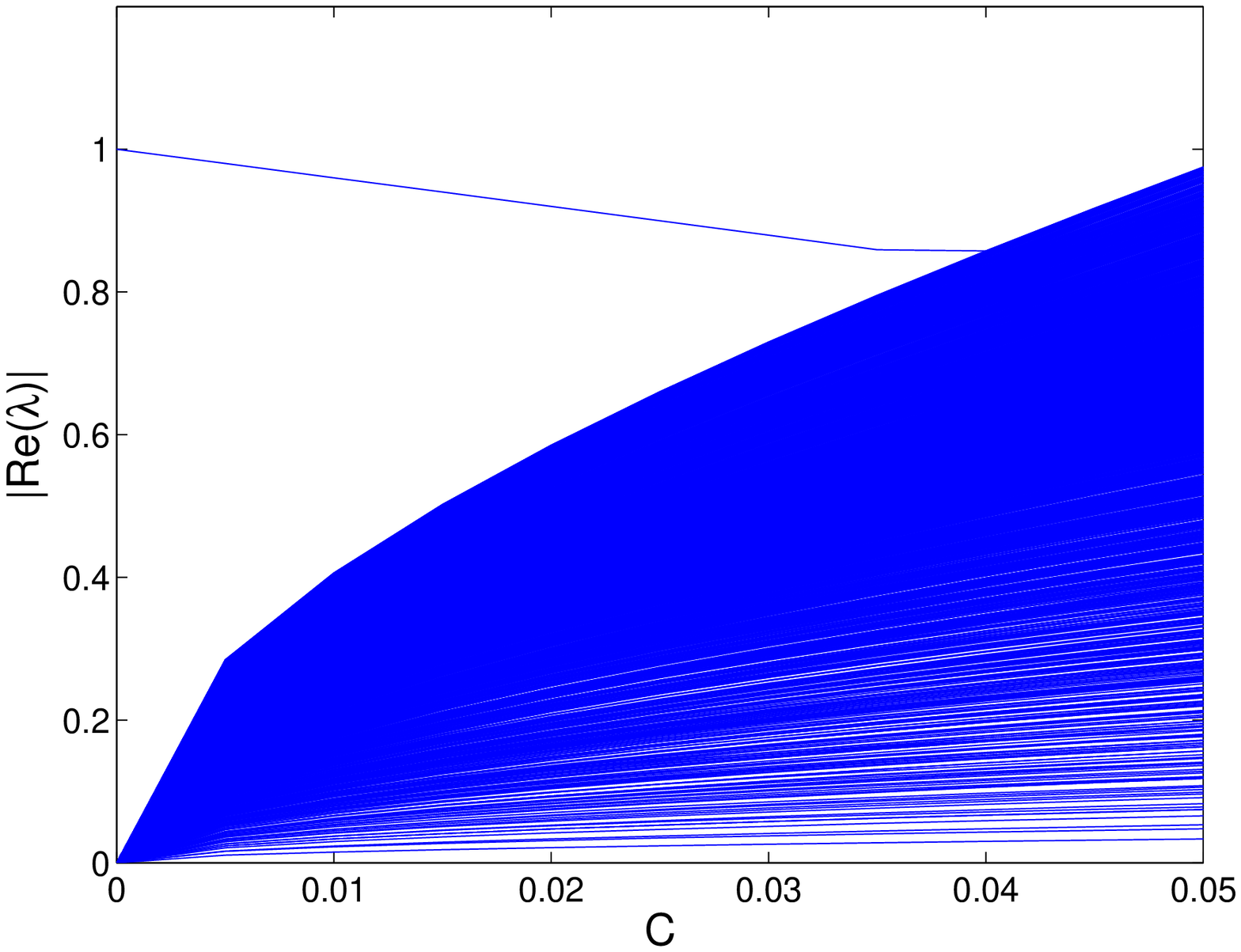} &
    \includegraphics[width=7cm]{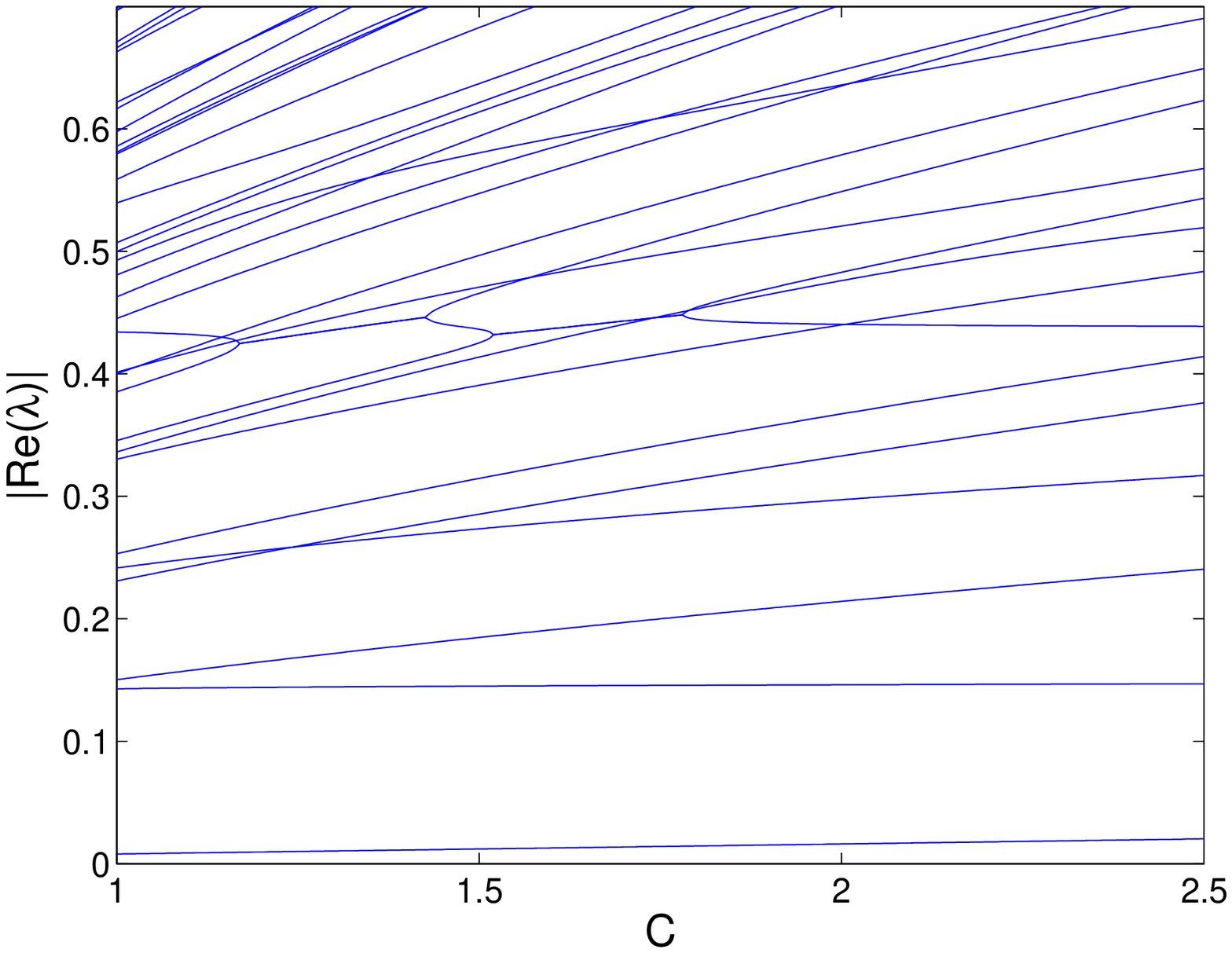}
\end{tabular}
\caption{Real part of the stability eigenfrequencies for $S=2$. The
panels show zooms of two different regions.} \label{fig:S2stab}
\end{center}
\end{figure}

\begin{figure}
\begin{center}
\begin{tabular}{cc}
    \includegraphics[width=7cm]{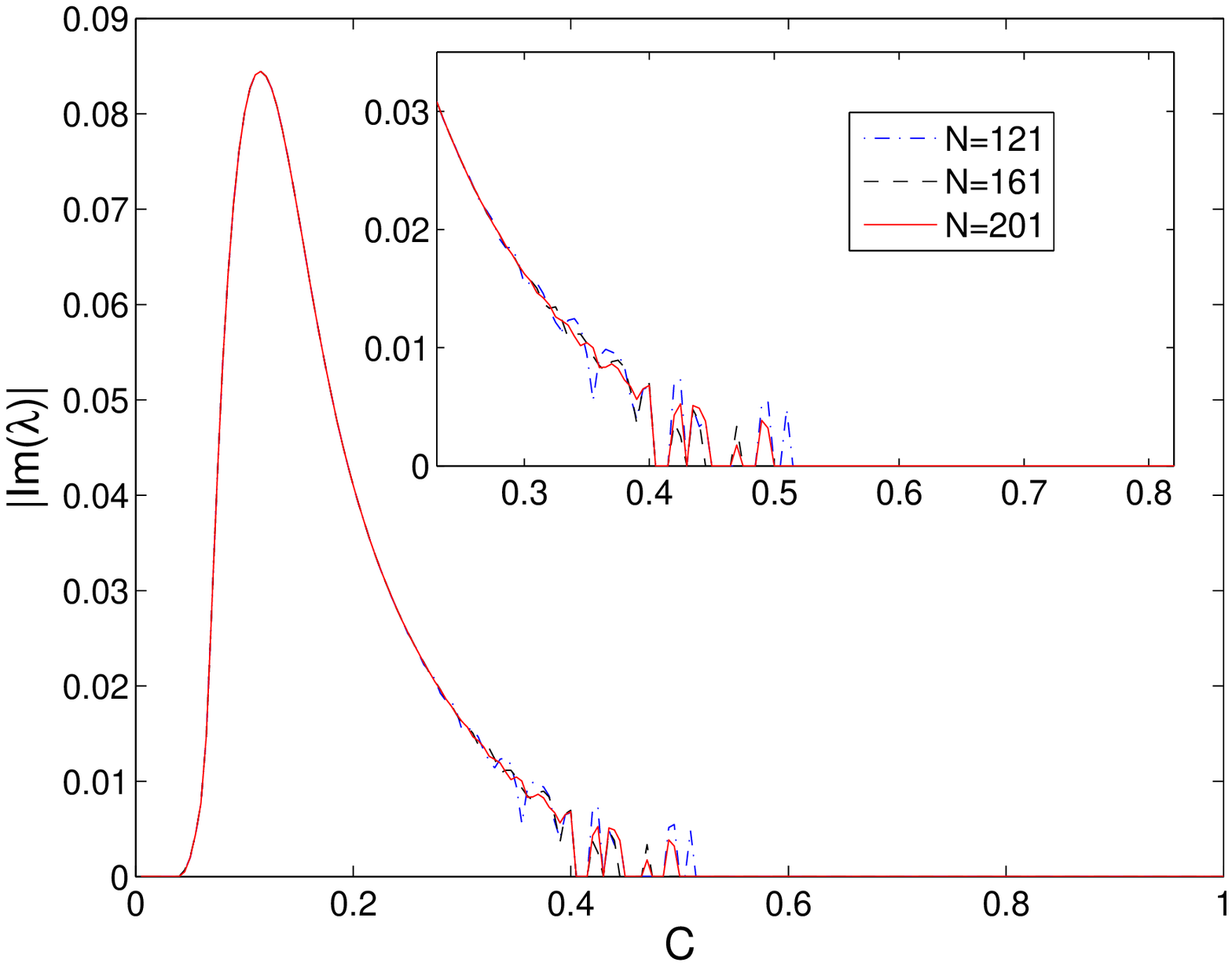} &
    \includegraphics[width=7cm]{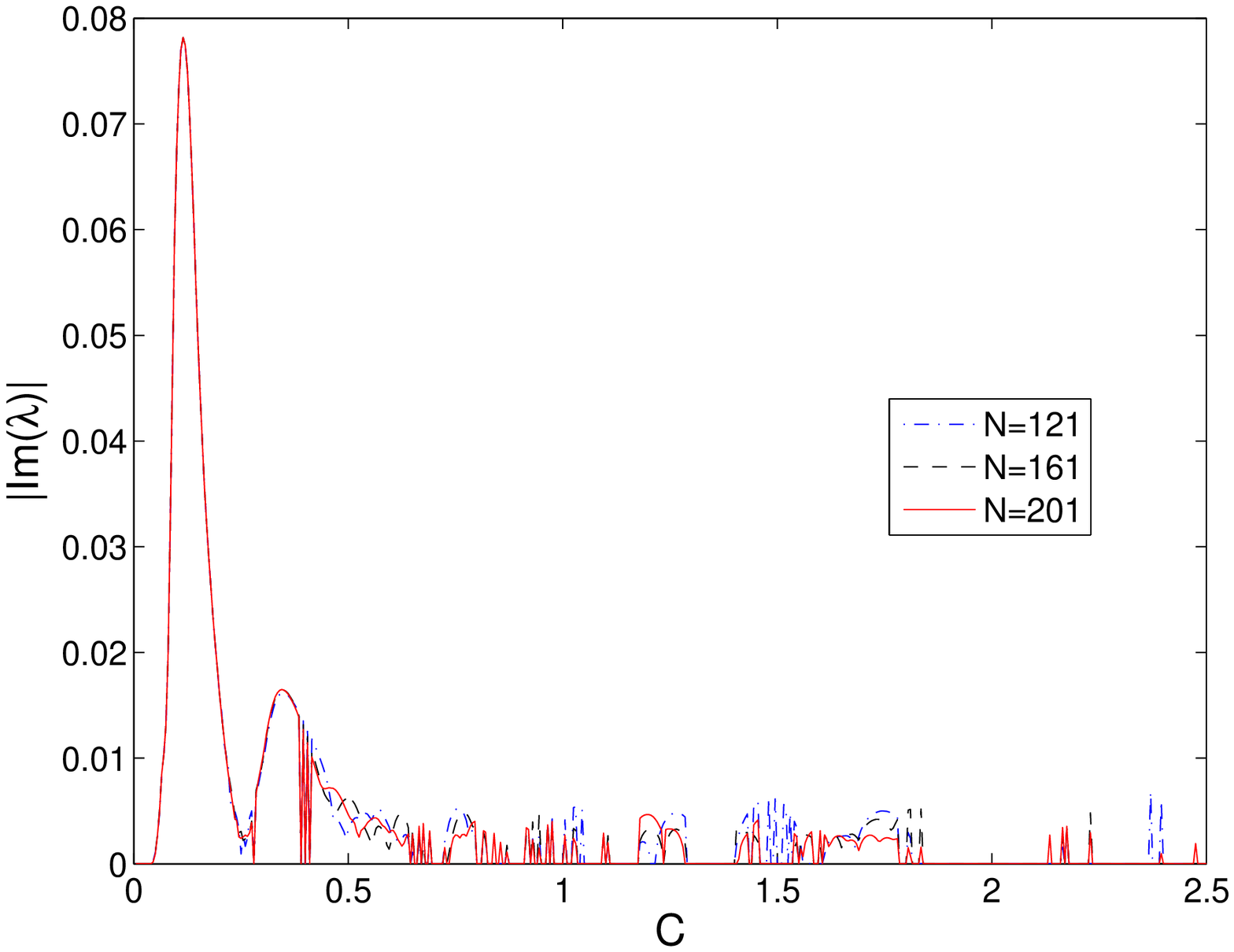}
\end{tabular}
\caption{Imaginary part of the stability eigenfrequencies for $S=1$ (left
panel) and $S=2$ (right panel), as a function of the coupling strength
$C$. This corresponds to the growth rate of the corresponding instability.} \label{fig:stab}
\end{center}
\end{figure}

On the other hand, by analogy to the one dimensional calculation, it
is straightforward to compute the dispersion relation characterizing
the eigenfrequencies of the continuous spectrum (using
$\{p_{n,m},q_{n,m}^{\star}\} = \{P,Q^{\star}\}\exp[i (k_n n + k_m
m)]$, deriving a $2 \times 2$ homogeneous linear system for $P$ and
$Q$ and demanding that its determinant be zero) as extending through
the interval $\lambda \in [-\sqrt{64 C^2 + 16 C}, \sqrt{64 C^2 + 16
C}]$. Therefore, the collision of the point spectrum (negative
energy) eigenvalue with the band edge of the continuous spectrum
yields a prediction for the critical point of $C_{cr} \approx (2
\sqrt{3}-3)/12 \approx 0.0387$ in good agreement with the
corresponding numerical result above. At $C=C_{cr}$ the system
experiences a Hamiltonian Hopf bifurcation. In consequence, there
exists an eigenvalue quartet
$\{\lambda,\lambda^*,-\lambda,-\lambda^*\}$. When $C$ increases, a
cascade of Hopf bifurcations takes place due to the interaction of a
localized mode with extended modes, as it was observed in
one-dimensional dark solitons \cite{johansson} (see also
\cite{MA98}, \cite{AACR02} to illustrate the appearance of this
phenomenon in Klein--Gordon lattices). This cascade implies the
existence of stability windows between inverse Hopf bifurcations and
direct Hopf bifurcations. For $S=1$ vortices, each one of the
bifurcations takes place for decreasing $|{\rm Re}(\lambda)|$ when
$C$ grows, and, in consequence, the bifurcations cease at a given
value of $C$, as $|{\rm Re}(\lambda)|$ of the localized mode is
smaller than that of the lowest extended mode frequency [however, in
the infinite domain limit, this eventual restabilization would not
take place but for the limit of $C \rightarrow \infty$]. This fact
is illustrated in Fig. \ref{fig:S1stab}. A similar plot for the case
of the $S=2$ vortex is shown in Fig. \ref{fig:S2stab}. When the
lattice size tends to infinity ($N\rightarrow\infty$), the linear
mode band extends from zero to infinity and becomes dense; thus,
these stabilization windows should disappear at this limit. To
illustrate this point, we have considered lattices of up to $201
\times 201$ sites for the $S=1$ and $S=2$ vortices and have shown
the growth rate of the corresponding instabilities in Fig.
\ref{fig:stab}. The maximum growth rate (i.e. the largest imaginary
part of the stability eigenfrequencies) takes place at
$C\approx0.115$ for $S=1$ and $S=2$ and being ${\rm
Im}(\lambda)\approx0.0845$ (0.0782) for $S=1$ ($S=2$).

\section{Harmonic Trap}

In this section, we consider the effect of introducing a harmonic
trap. Thus, Eq. (\ref{eq:stat}) is modified:

\begin{equation}\label{eq:stattrap}
C\, \Delta \phi_{n,m} + (1-|\phi_{n,m}|^2-V_{n,m}) \phi_{n,m}=0,
\end{equation}
with the parabolic potential of the form\footnote{The factor
$1/C$ appears when discretizing
the continuum equation given that $C=1/h^2$,  where
$h$ the lattice spacing. In particular, we have used
$r=\sqrt{x^2+y^2}=h\sqrt{m^2+n^2}=hr_{n,m}=r_{n,m}/\sqrt{C}$.}
\begin{equation}
V_{n,m}=\frac{1}{C}\Omega^2r_{n,m}^2\ .
\end{equation}

\begin{figure}
\begin{center}
\begin{tabular}{cc}
    \includegraphics[width=7cm]{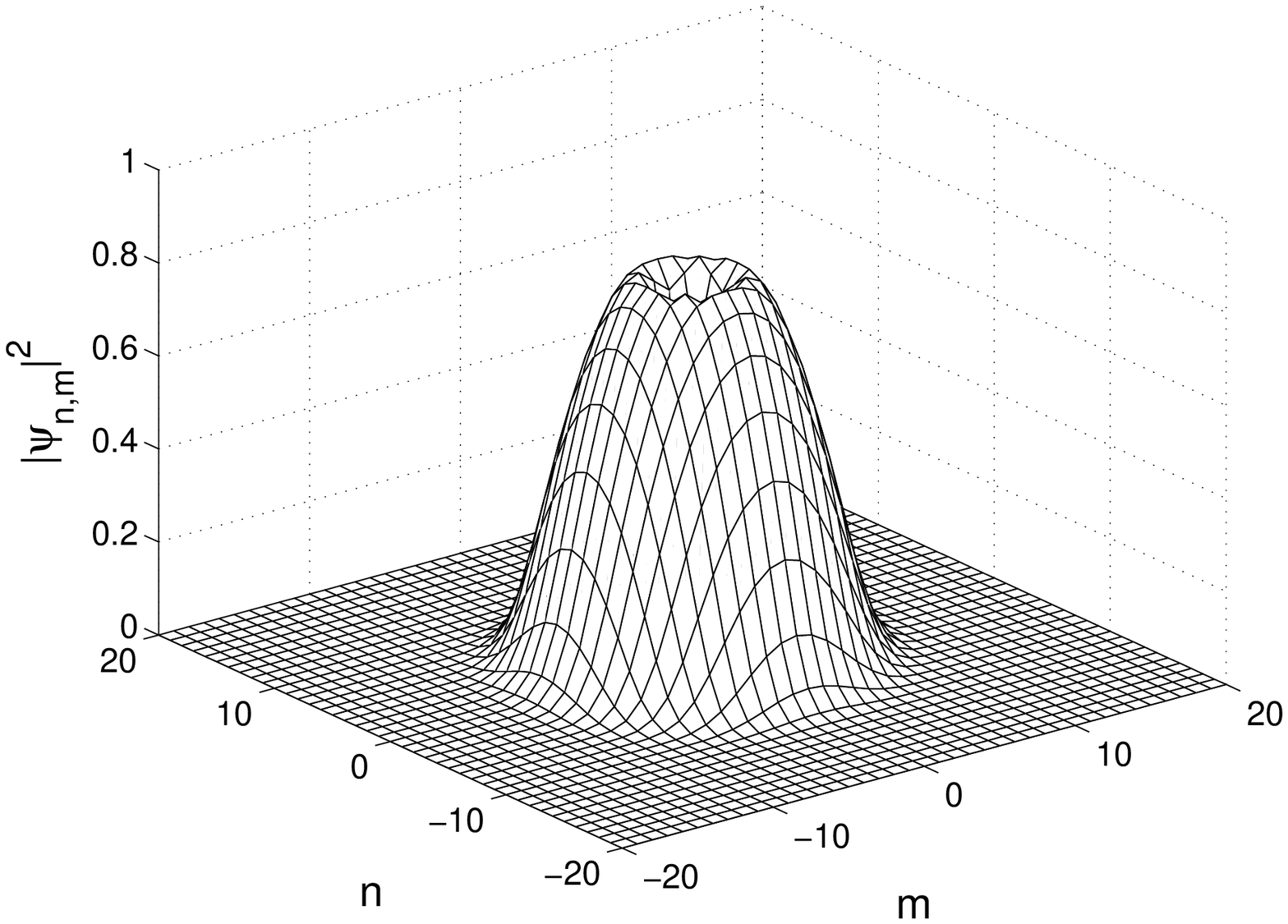} &
    \includegraphics[width=7cm]{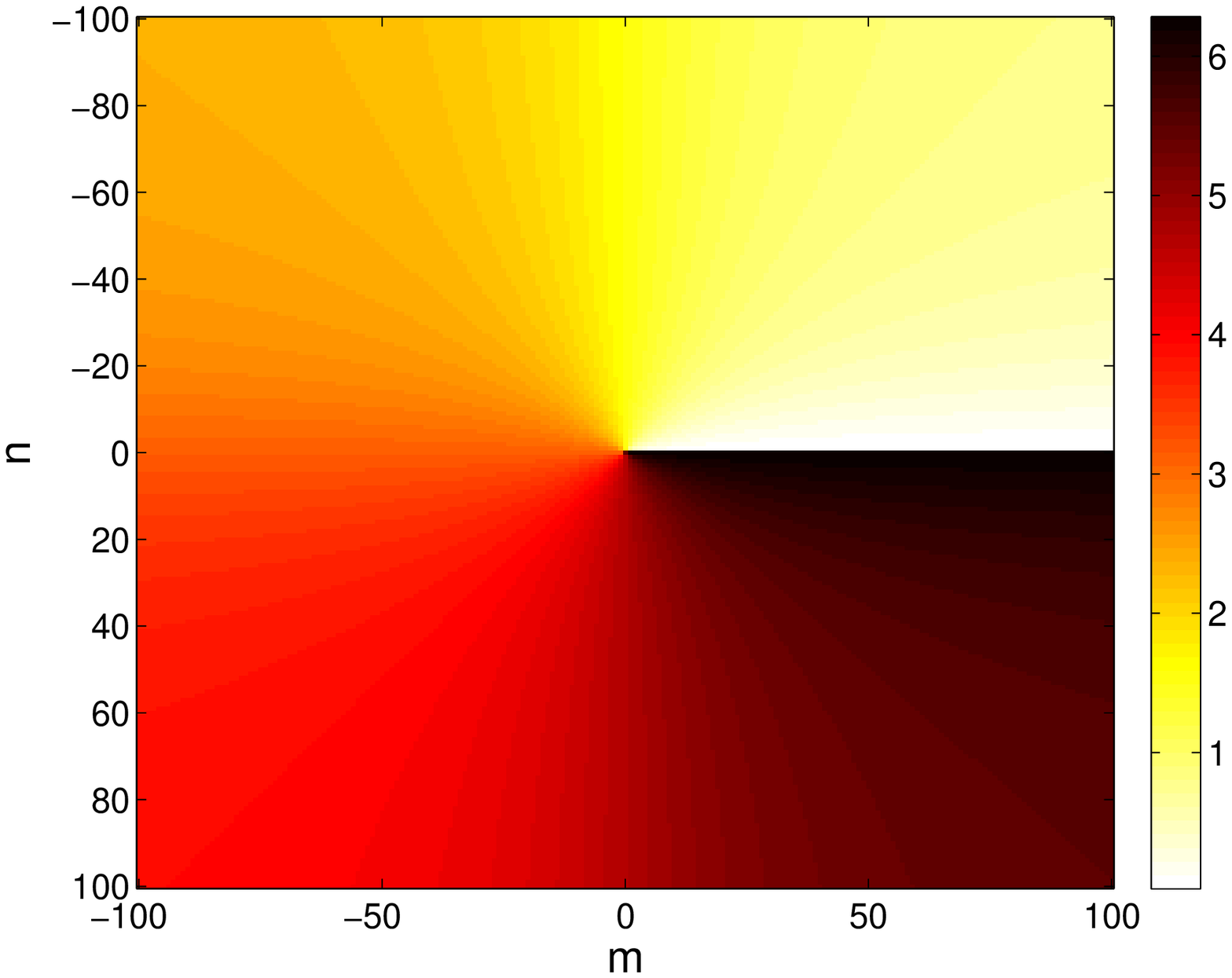}
\end{tabular}
\caption{Vortex soliton with $S=1$ and $C=0.5$ in a harmonic trap
with $\Omega=0.1$. (Left panel) density Profile; (Right panel)
angular dependence.} \label{fig:TrapS1}
\end{center}
\end{figure}

Fig. \ref{fig:TrapS1}
%and \ref{fig:TrapS2}
shows a typical  example of such a
discrete vortex structure in the presence of an external trapping
potential. The method presented in Section \ref{num_method} is again
used and converges unhindered by the presence of the magnetic trap.
Notice that to include the trapping effect of the potential, we only
modify the initial guess proposed in Section \ref{num_method}
through multiplying it by the so-called Thomas-Fermi profile of
$\sqrt{\max(0,1-V_{n,m})}$ \cite{pethick,stringari}; the resulting 
guess converges
even for small values of $C$ (such as the one used in Fig. \ref{fig:TrapS1}). 
These vortices can be
continued up to $C\rightarrow\infty$ and will converge to the
corresponding continuum trapped vortices (for a recent discussion of
such vortices in the presence of external potentials see e.g.
\cite{klaw}).

\begin{figure}
\begin{center}
\begin{tabular}{cc}
    \includegraphics[width=7cm]{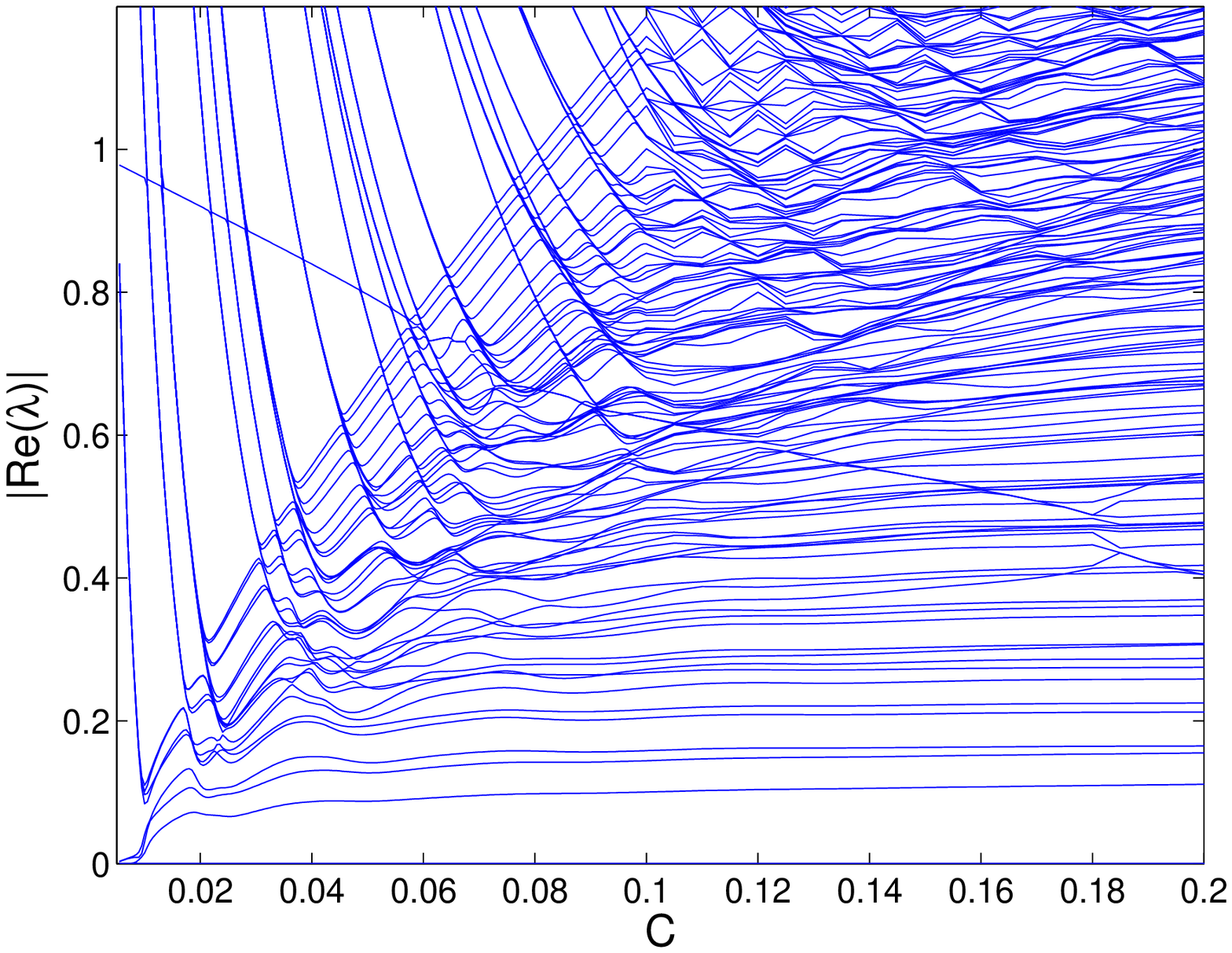} &
    \includegraphics[width=7cm]{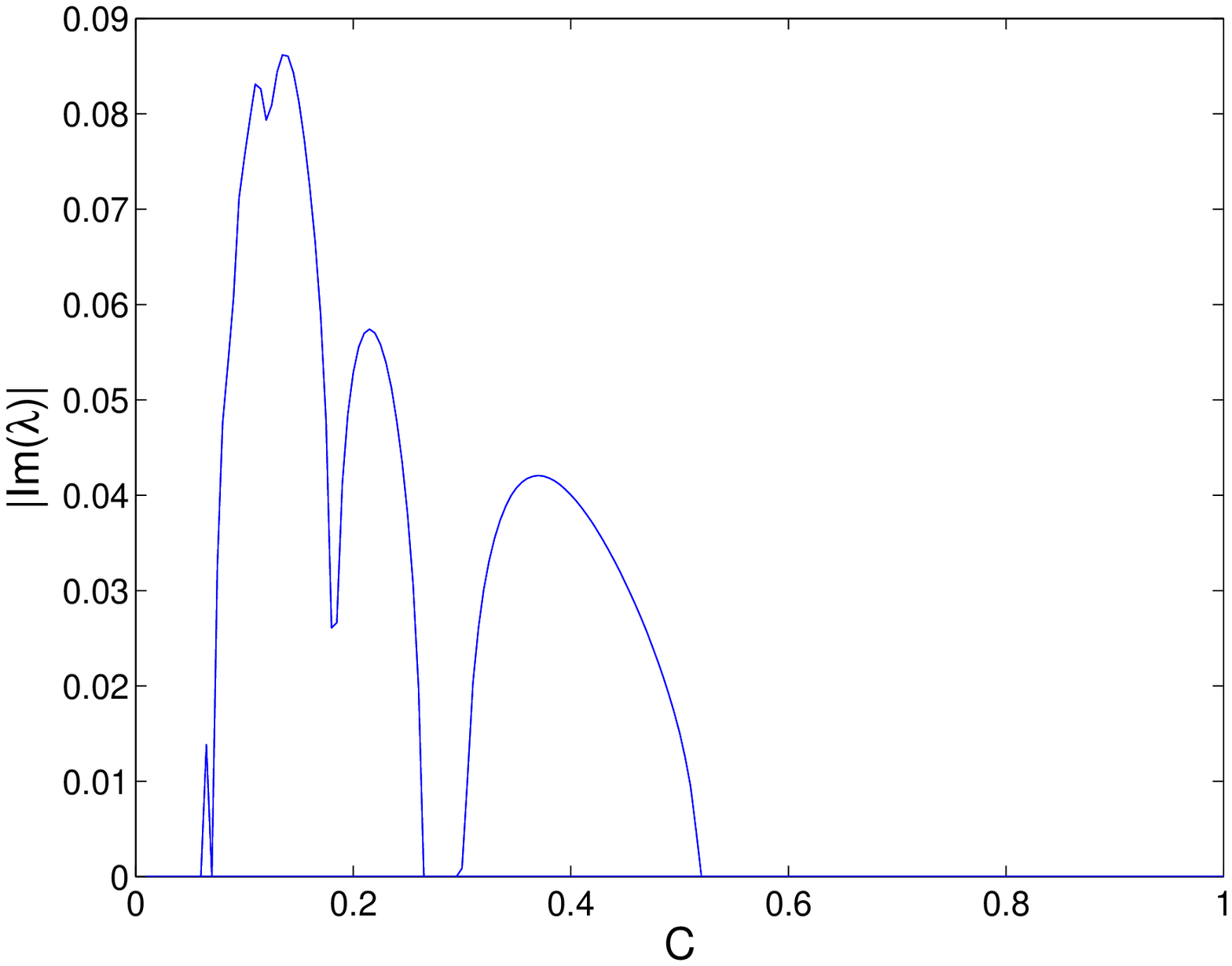}
\end{tabular}
\caption{Real part (left panel) and imaginary part (right panel) of
the stability eigenfrequencies for $S=1$ as a function of the
coupling strength $C$ and a harmonic trap with $\Omega=0.1$.}
\label{fig:stabMT1}
\end{center}
\end{figure}

\begin{figure}
\begin{center}
\begin{tabular}{cc}
    \includegraphics[width=7cm]{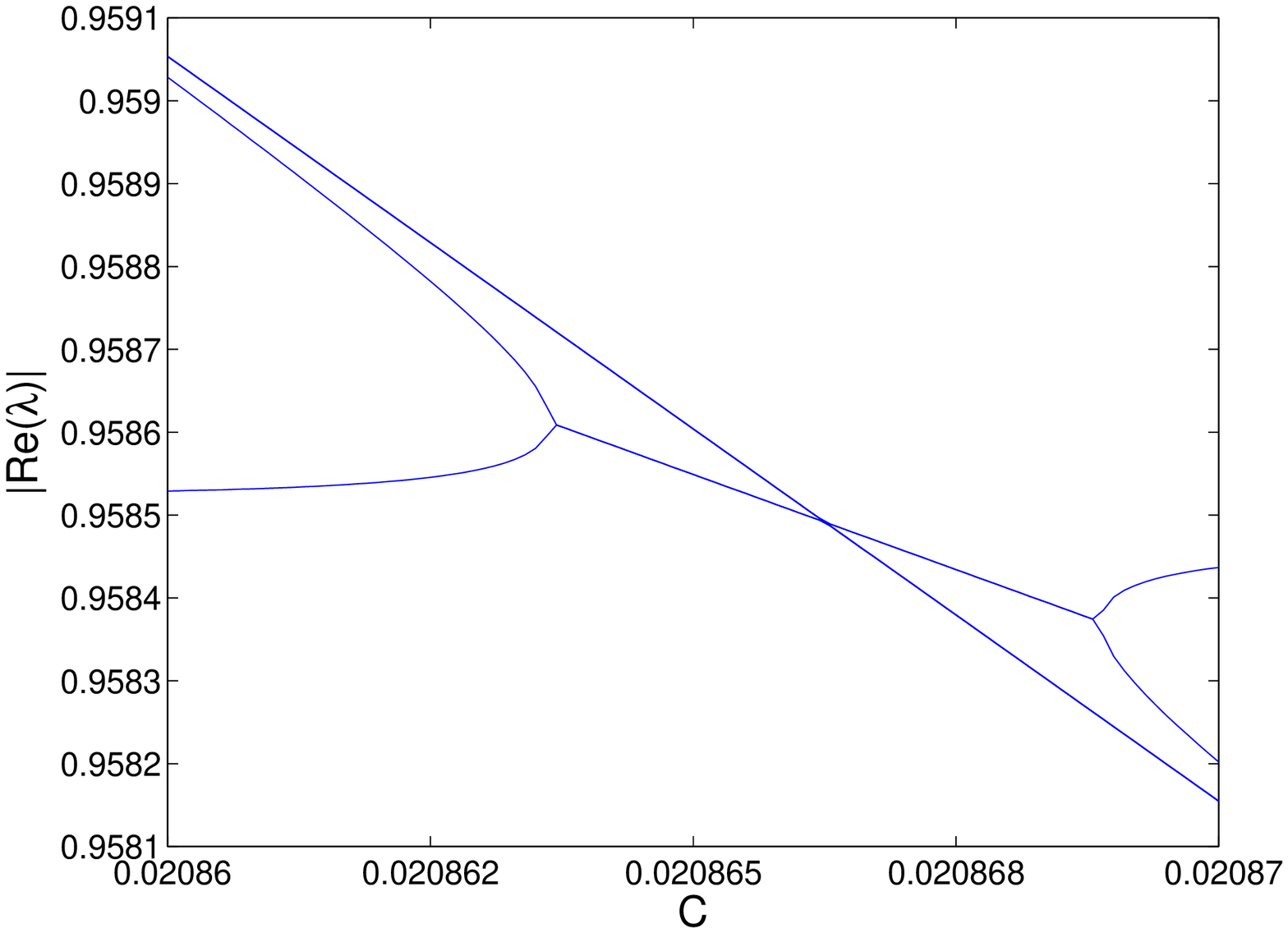} &
    \includegraphics[width=7cm]{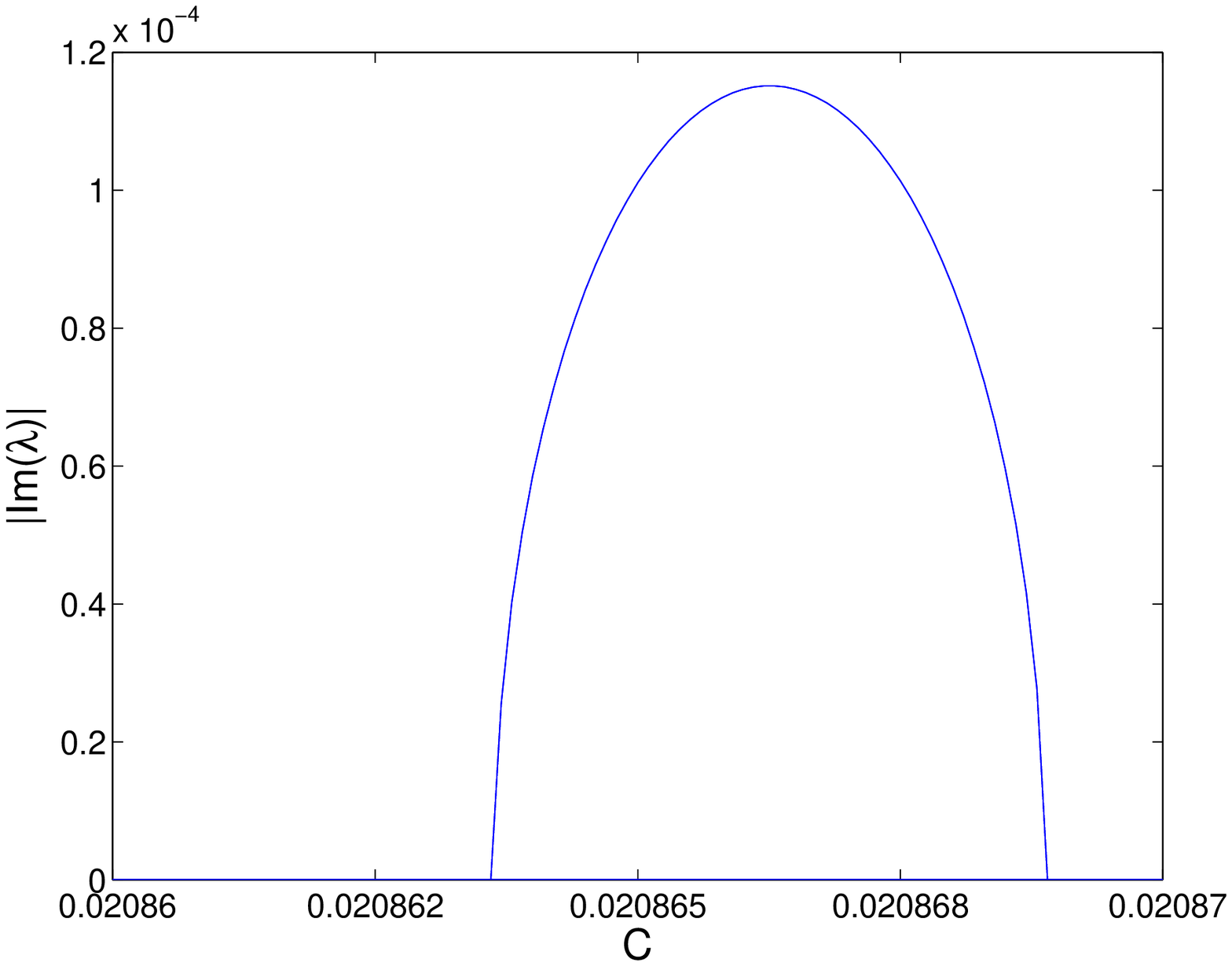}
\end{tabular}
\caption{Same as Fig. \ref{fig:stabMT1} with a zoom around the first
bifurcation. From this figure, it is clear that there is a Hopf
bifurcation that destabilizes the vortex and an inverse Hopf. This
pair of bifurcations takes place in a small interval of $C$ with a
length of approximately $3\times10^{-6}$.} \label{fig:stabMT2}
\end{center}
\end{figure}

The stability of such structures is also examined in Figs.
\ref{fig:stabMT1} and \ref{fig:stabMT2}. The sole type of
instability observed is an oscillatory one, with alternating windows
of destabilization and restabilization. However, since the harmonic
trap is well-known \cite{pethick,stringari} to discretize the
spectrum of excitations, these windows of
instability/restabilization are ``true'' ones (due to collisions of
the ``negative energy'' mode of the vortex with the point spectrum
of the background), rather than artificial ones (caused by the
finite size of the computational domain). In fact, in this case, the maximum
imaginary part of the eigenvalues does not depend on the number of
grid points used (provided that the domain ``encompasses'' the 
harmonically trapped vortex). 
For high enough $C$, the charge $S=1$ vortex is always
found to stabilize
%, while for $S=2$, this depends on the chemical
%potential
\cite{pu,klaw}. It is interesting to also note that although the fundamental
destabilization scenario indicated by the right panel of Fig. \ref{fig:stabMT1}
has very strong parallels with its untrapped analog, the left panel
of the figure indicates multiple additional collisions for smaller values
of $C$. The negative Krein sign of the translational eigenvalue, 
discussed previously, suggests that these collisions should also result
in oscillatory instabilities, although this is not discernible in the
left column of Fig. \ref{fig:stabMT1}. A relevant clarification to this
apparent paradox is provided by Fig. \ref{fig:stabMT2} which clearly
illustrates that the oscillatory instabilities do indeed arise but,
in fact, emerge and disappear (the latter through inverse Hopf bifurcations)
over very tiny parametric intervals of $C$ (and are, thus, apparently
invisible over the scale of Fig. \ref{fig:stabMT1}).

%\begin{figure}
%\begin{center}
%\begin{tabular}{cc}
%    \includegraphics[width=7cm]{MTS2a.eps} &
%    \includegraphics[width=7cm]{MTS2b.eps}
%\end{tabular}
%\caption{Vortex soliton with $S=2$ and $C=0.5$ in a harmonic trap
%with $\Omega=0.1$. (Left panel) density Profile; (Right panel)
%angular dependence.} \label{fig:TrapS2}
%\end{center}
%\end{figure}

%\begin{figure}
%\begin{center}
%\begin{tabular}{cc}
%    \includegraphics[width=7cm]{MTS1real.eps} &
%    \includegraphics[width=7cm]{MTS2real.eps}
%\end{tabular}
%\caption{Real part of the stability eigenfrequencies for $S=1$ (left
%panel) and $S=2$ (right panel), as a function of the coupling
%strength $C$. In both cases, there is a harmonic trap with
%$\Omega=0.1$.} \label{fig:stabMT1}
%\end{center}
%\end{figure}

%\begin{figure}
%\begin{center}
%\begin{tabular}{cc}
%    \includegraphics[width=7cm]{MTS1imag.eps} &
%    \includegraphics[width=7cm]{MTS2imag.eps}
%\end{tabular}
%\caption{Imaginary part of the stability eigenfrequencies for $S=1$
%(left panel) and $S=2$ (right panel), as a function of the coupling
%strength $C$. In both cases, there is a harmonic trap with
%$\Omega=0.1$.} \label{fig:stabMT2}
%\end{center}
%\end{figure}

\section{Conclusions and Future Directions}

In the present paper, we examined the discrete analog of continuum
defocusing vortices which are perhaps the prototypical coherent
structure in the two-dimensional nonlinear Schr{\"o}dinger equation.
We illustrated how to systematically obtain such structures through
an appropriate continuation of the amplitude and phase profiles from
the anti-continuum limit, and also discussed how to perform such a
continuation from the continuum limit (at least for single core
vortices). Such a continuation as a function of the coupling
strength revealed significant analogies between these defocusing
discrete vortices and their 1d analog of the discrete dark solitons,
which are stable from coupling $C=0$ up to a critical coupling and
are subsequently unstable for all higher couplings up to $C
\rightarrow \infty$ (when they become restabilized). Something
similar was observed and quantified in the case of discrete
vortices. In addition to the most fundamental structures of
topological charge $S=1$, structures of higher charge such as $S=2$
were obtained by similar means.

A natural topic for a more detailed future study arising from
the present work concerns the understanding of multi-vortex
bound states and their stability properties, as well as their
detailed continuation as a function of the coupling and eventual
disappearance as the coupling becomes sufficiently large.
Another possible direction would be to examine such defocusing
vortices in multi-component models (in analogy e.g., to the
bright discrete vortices of \cite{pelinovsky2d}; see also references
therein). There it would be of interest to study the similarities
and differences of bound states of the same charge versus ones
of, say, opposite charges. For these more demanding computations
(as well as possibly ones associated with the 3d version of the
present model \cite{our3d}), more intensive numerical computations
will be needed which may be aided by virtue of parallel implementation
\cite{Faustino}.
Such studies are currently in progress and will be reported
in a future publication.

\section*{Acknowledgments}

We acknowledge Faustino Palmero for his help with the implementation
of the numerical routines. PGK gratefully acknowledges support from
NSF-CAREER, NSF-DMS-0505663, NSF-0619492 and from the Alexander
von Humboldt Foundation through a Research Fellowship.


\begin{thebibliography}{99}

\bibitem{pethick} C.J. Pethick and H. Smith,
{\it Bose-Einstein condensation in dilute gases}, Cambridge University
Press (Cambridge, 2002).


\bibitem{stringari} L.P. Pitaevskii and S. Stringari,
{\it Bose-Einstein Condensation}, Oxford University Press (Oxford, 2003).

\bibitem{vort1} M. R. Matthews, B. P. Anderson, P. C. Haljan, D. S. Hall, C. E. Wieman, and E. A. Cornell,
\newblock Vortices in a Bose-Einstein condensate,
\newblock  Physical Review Letters {\bf 83}, 2498-2501 (1999).

\bibitem{vort2} K. W. Madison, F. Chevy, W. Wohlleben, and J. Dalibard,
\newblock Vortex Formation in a stirred Bose-Einstein condensate,
\newblock Physical Review Letters {\bf 84}, 806-809 (1999).

\bibitem{vort3} S. Inouye, S. Gupta, T. Rosenband, A. P. Chikkatur,
A. G{\"o}rlitz, T. L. Gustavson, A. E. Leanhardt, D. E. Pritchard, and
W. Ketterle,
\newblock Observation of vortex phase singularities in Bose-Einstein
condensates,
\newblock Physical Review Letters {\bf 87}, 080402, 4 pages (2001).

\bibitem{latt1} J.R. Abo-Shaeer,  C. Raman, J.M. Vogels, W. Ketterle,
\newblock Observation of vortex lattices in Bose-Einstein condensates
\newblock Science {\bf 292}, 476-479 (2001).

\bibitem{latt2} J.R. Abo-Shaeer, C. Raman, W. Ketterle,
\newblock Formation and decay of vortex lattices in Bose-Einstein
condensates at finite temperatures
\newblock Physical Review Letters {\bf 88}, 070409, 4 pages (2002).

\bibitem{latt3} P. Engels, I. Coddington, P.C. Haljan and
E.A. Cornell,
Nonequilibrium effects of anisotropic compression applied
to vortex lattices in Bose-Einstein condensates,
\newblock Physical Review Letters {\bf 89}, 100403, 4 pages
(2002).

\bibitem{pismen} L.M. Pismen, {\it Vortices in nonlinear fields},
Oxford University Press (Oxford, 1999)

\bibitem{fetter} A.L. Fetter and A.A. Svidzinsky,
J. Phys. Condens. Matter {\bf 13}, R135 (2001).

\bibitem{our} P.G. Kevrekidis, R. Carretero-Gonz{\'a}lez,
D.J. Frantzeskakis and I.G. Kevrekidis, Mod. Phys. Lett. B {\bf 18}, 1481
(2004).

\bibitem{book} P.G. Kevrekidis, D.J. Frantzeskakis and
R. Carretero-Gonz\'alez (Eds.),
{\it Emergent Nonlinear Phenomena in Bose-Einstein Condensates},
Springer-Verlag (Heidelberg, 2008).

\bibitem{berloff1} N.G. Berloff,
{\it Quantum vortices, travelling coherent structures and superfluid
turbulence}, preprint available at:
http://www.damtp.cam.ac.uk/user/ngb23.

\bibitem{ournew} R. Carretero-Gonz\'alez, D.J. Frantzeskakis and
P.G. Kevrekidis, {\it Nonlinear waves in Bose-Einstein condensates:
physical relevance and mathematical techniques},
preprint available at:
http://www-rohan.sdsu.edu/$\sim$rcarrete/

\bibitem{konotop} V.A. Brazhnyi and V.V. Konotop,
Mod. Phys. Lett. B {\bf 18}, 627 (2004).

\bibitem{morsch} O. Morsch and M. Oberthaler,
Rev. Mod. Phys. {\bf 78}, 179 (2006).

\bibitem{jpb} P.G. Kevrekidis, R. Carretero-Gonz{\'a}lez, G. Theocharis,
D.J. Frantzeskakis and B.A. Malomed,
J. Phys. B {\bf 36}, 3467 (2003).

\bibitem{TS} A. Trombettoni and A. Smerzi,
Phys. Rev. Lett. {\bf 86}, 2353 (2001).

\bibitem{konotop2} F.Kh. Abdullaev, B.B. Baizakov, S.A. Darmanyan,
V.V. Konotop and M. Salerno, Phys. Rev. A {\bf 64}, 043606 (2001).

\bibitem{us} G.L. Alfimov, P.G. Kevrekidis, V.V. Konotop and M. Salerno,
Phys. Rev. E {\bf 66}, 046608 (2002).

\bibitem{usnew}  F.Kh. Abdullaev, Yu.V. Bludov, S.V. Dmitriev, P.G.
Kevrekidis, and V. V. Konotop, Phys. Rev. E {\bf 77}, 016604 (2008)

\bibitem{lederer1} T. Pertsch, U. Peschel, F. Lederer, J. Burghoff,
M. Will, S. Nolte and A.  T{\"u}nnermann, Opt. Lett. {\bf 29}, 468 (2004).

\bibitem{lederer2} A. Szameit, J. Burghoff, T. Pertsch, S. Nolte,
A T{\"u}nnermann, Opt. Express {\bf 14}, 6055 (2006).


\bibitem{lederer3}  A. Szameit, Y. Kartashov, F. Dreisow, M. Heinrich,
V.A. Vysloukh, T. Pertsch, S. Nolte, A. T{\"u}nnermann, F. Lederer
and L. Torner,  arXiv:0802.3196.

\bibitem{kip} E. Smirnov, C.E. R{\"u}ter, M. Stepi{\'c},
D. Kip and V. Shandarov, Phys. Rev. E {\bf 74}, 065601 (R).

\bibitem{johansson} M Johansson and YuS Kivshar.
Phys. Rev. Lett. {\bf 82}, 85 (1999).


\bibitem{fitrakis} E.P. Fitrakis, P.G. Kevrekidis,
H. Susanto and D.J. Frantzeskakis,
Phys. Rev. E {\bf 75}, 066608 (2007).

\bibitem{Berloff} NG Berloff. J. Phys. A: Math. Gen. {\bf 37}, 1617
(2004).

\bibitem{MA94} RS MacKay and S Aubry. Nonlinearity {\bf 7}, 1623
(1994).

\bibitem{MHM07} A Maluckov, L Had\v{z}ievski and BA Malomed. Phys. Rev.
E {\bf 76}, 046605 (2007).

\bibitem{giorgo} G. Theocharis, P.G. Kevrekidis, M.K. Oberthaler, and
D.J. Frantzeskakis,   Phys. Rev. A {\bf 76}, 045601 (2007).

\bibitem{pu} H. Pu, C.K. Law, J.H. Eberly, and N. P. Bigelow,Phys. Rev. A
{\bf 59}, 1533 (1999); Y. Kawaguchi and T. Ohmi, Phys. Rev. A
{\bf 70}, 043610 (2004); J.A.M. Huhtam{\"a}ki, M. M{\"o}t{\"o}nen and
S.M.M. Virtanen, Phys. Rev. A {\bf 74}, 063619 (2006).

\bibitem{MA98} JL Mar\'{\i}n and S Aubry. Physica D {\bf 119}, 163 (1998).

\bibitem{AACR02} A \'{A}lvarez, JFR Archilla, J Cuevas and FR Romero.
New J. Phys. {\bf 4}, 72 (2002).


%\bibitem{Kalo} G Kalosakas. Physica D {\bf 216}, 44 (2006).

%\bibitem{kalo1} H.
%Susanto and M Johansson, Phys. Rev. E {\bf 72}, 016605 (2005).

\bibitem{klaw} K.J.H. Law, L. Qiao, P.G. Kevrekidis and I.G. Kevrekidis,
Phys. Rev. A {\bf 77}, 053612 (2008).

%\bibitem{pu}
%H. Pu, C. K. Law, J. H. Eberly, and N. P. Bigelow, Phys. Rev. A
%{\bf 59}, 1533 (1999).

\bibitem{pelinovsky2d} P.G. Kevrekidis and D.E. Pelinovsky,
Proc. Roy. Soc. A {\bf 462}, 2671 (2006).

\bibitem{our3d} P.G. Kevrekidis, B.A. Malomed, D.J. Frantzeskakis
and R. Carretero-Gonz\'alez, Phys. Rev. Lett. {\bf 93}, 080403 (2004).
R. Carretero-Gonz\'alez, P.G. Kevrekidis, B.A. Malomed and
D.J. Frantzeskakis, Phys. Rev. Lett. {\bf 94}, 203901 (2005).

\bibitem{Faustino} F Palmero. Breathers in Quantum Lattices.
In P Alberigo, G Erbacci and F Garofalo, Eds. Science and
Supercomputing in Europe (report 2005) , pages 758-762. CINECA,
Bologna (Italy), 2006.

\end{thebibliography}
\end{document}